\documentclass[fleqn,usenatbib,useAMS]{mnras}

\usepackage{newtxtext,newtxmath}
\usepackage[T1]{fontenc}
\usepackage[normalem]{ulem}
\usepackage[dvipsnames]{xcolor}

\DeclareRobustCommand{\VAN}[3]{#2}
\let\VANthebibliography\thebibliography
\def\thebibliography{\DeclareRobustCommand{\VAN}[3]{##3}\VANthebibliography}

\usepackage{graphicx}	% Including figure files
\usepackage{amsmath}	% Advanced maths commands
\usepackage{amssymb}	% Extra maths symbols
\usepackage{scalerel,tikz}
\usetikzlibrary{svg.path}
\definecolor{orcidlogocol}{HTML}{A6CE39}
\tikzset{orcidlogo/.pic={
		\fill[orcidlogocol] svg{M256,128c0,70.7-57.3,128-128,128C57.3,256,0,198.7,0,128C0,57.3,57.3,0,128,0C198.7,0,256,57.3,256,128z};
		\fill[white] svg{M86.3,186.2H70.9V79.1h15.4v48.4V186.2z}
		svg{M108.9,79.1h41.6c39.6,0,57,28.3,57,53.6c0,27.5-21.5,53.6-56.8,53.6h-41.8V79.1z M124.3,172.4h24.5c34.9,0,42.9-26.5,42.9-39.7c0-21.5-13.7-39.7-43.7-39.7h-23.7V172.4z}
		svg{M88.7,56.8c0,5.5-4.5,10.1-10.1,10.1c-5.6,0-10.1-4.6-10.1-10.1c0-5.6,4.5-10.1,10.1-10.1C84.2,46.7,88.7,51.3,88.7,56.8z};
}}
\newcommand\orcidicon[1]{\href{https://orcid.org/#1}{\mbox{\scalerel*{
				\begin{tikzpicture}[yscale=-1,transform shape]
					\pic{orcidlogo};
				\end{tikzpicture}
			}{|}}}}

\title[satellite galaxy-galaxy lensing]{Assessing Mass Loss and Stellar-to-Halo Mass Ratio of Satellite Galaxies: A Galaxy-Galaxy Lensing Approach Utilizing DECaLS DR8 Data}

\author[Chunxiang Wang et al.]{
	Chunxiang Wang$^{1,2,3}$\thanks{E-mail:\url{chunxiang_wang@sina.cn}},
	Ran Li$^{1,2,3}$\thanks{E-mail: \url{ranl@bao.ac.cn}},
	Huanyuan Shan$^{4,5,6}$,
	Weiwei Xu\orcidicon{0000-0002-9587-6683}$^{1,2,3,7}$,
    Ji Yao$^{4}$,
    Yingjie Jing$^{1,2}$,\and
	Liang Gao$^{1,2,3,8}$,
	Nan Li\orcidicon{0000-0001-6800-7389}$^{1,9}$,
	Yushan Xie$^{3,4}$,
    Kai Zhu\orcidicon{0000-0002-2583-2669}$^{1,2,3}$,
    Hang Yang\orcidicon{0000-0003-3279-0134}$^{1,2,3}$,
    Qingze Chen$^{1,2,3}$
	\\
	$^{1}$National Astronomical Observatories, Chinese Academy of Sciences, Beijing 100101, China\\
	$^{2}$Institute for Frontiers in Astronomy and Astrophysics, Beijing Normal University, Beijing 102206, China\\
	$^{3}$School of Astronomy and Space Science, University of Chinese Academy of Sciences, Beijing 100049, China\\
    $^{4}$Shanghai Astronomical Observatory (SHAO), Nandan Road 80, Shanghai 200030, China\\
    $^{5}$Key Laboratory of Radio Astronomy and Technology, Chinese Academy of Sciences, A20 Datun Road, Chaoyang District, Beijing, 100101, P. R. China\\
    $^{6}$University of Chinese Academy of Sciences, Beijing 100049, China\\
	$^{7}$The Kavli Institute for Astronomy and Astrophysics, Peking University (KIAA-PKU), Beijing, China\\
	$^{8}$Institute for Computational Cosmology, Department of Physics, University of Durham, South Road, Durham, DH1 3LE, UK\\
	$^{9}$Key lab of Space Astronomy and Technology, National Astronomical Observatories, 20A Datun Road, Chaoyang District, Beijing 100012, China\\
    }

\date{Accepted XXX. Received YYY; in original form ZZZ}

\pubyear{2023}

\begin{document}
\label{firstpage}
\pagerange{\pageref{firstpage}--\pageref{lastpage}}
\maketitle

% Abstract of the paper
\begin{abstract}
The galaxy-galaxy lensing technique allows us to measure the subhalo mass of satellite galaxies, studying their mass loss and evolution within galaxy clusters and providing direct observational validation for theories of galaxy formation. In this study, we use the weak gravitational lensing observations from DECaLS DR8, in combination with the redMaPPer galaxy cluster catalog from Sloan Digital Sky Survey data (SDSS) DR8 to accurately measure the dark matter halo mass of satellite galaxies. We confirm a significant increase in the stellar-to-halo mass ratio of satellite galaxies with their halo-centric radius, indicating clear evidence of mass loss due to tidal stripping. Additionally, we find that this mass loss is strongly dependent on the mass of the satellite galaxies, with satellite galaxies above $10^{11}~{\rm M_{\odot}/h}$ experiencing more pronounced mass loss compared to lower mass satellites, reaching 86\% at projected halo-centric radius $0.5R_{\rm 200c}$. The average mass loss rate, when not considering halo-centric radius, displays a U-shaped variation with stellar mass, with galaxies of approximately $4\times10^{10}~{\rm  M_{\odot}/h}$ exhibiting the least mass loss, around 60\%. We compare our results with state-of-the-art hydrodynamical numerical simulations and find that the satellite galaxy stellar-to-halo mass ratio in the outskirts of galaxy clusters is higher compared to the predictions of the Illustris-TNG project about factor 5. Furthermore, the Illustris-TNG project's numerical simulations did not predict the observed dependence of satellite galaxy mass loss rate on satellite galaxy mass.
\end{abstract}

% Select between one and six entries from the list of approved keywords.
% Don't make up new ones.
%%%%%%%%%%%%%%%%%%%%%%%%%%%%%%%%%%%%%%%%%%%%%%%%%%

%%%%%%%%%%%%%%%%% BODY OF PAPER %%%%%%%%%%%%%%%%%%

\section{Introduction}

In the framework of modern cold dark matter cosmology, dark matter halos form hierarchically. In the early universe, the first to form are small dark matter halos, which grow into larger ones by merging and accreting matter \citep{Frenk_and_White_2012}. Gas collapses and condenses in the centers of dark matter halos, igniting stars and forming galaxies. Galaxies also evolve together with dark matter halos. When a small halo falls into a larger one, it experiences dynamical friction, tidal stripping, and tidal heating effects, gradually losing mass and eventually disintegrating\citep[e.g.][]{Gao2004, Springel2008, Gao_2012_Phoenix_MNRAS.425.2169G, xielizhi_2015MNRAS.454.1697X, Han2016, Niemiec_2019MNRAS.487..653N, Niemiec2022MNRAS.512.6021N}. In this process, galaxies transform into satellite galaxies within larger haloes, and their gas is removed through tidal stripping and ram pressure stripping, leading to the quenching of star formation\citep[e.g.][]{Wanglan2007, Guo2011, Wetzel2014}. Investigating the co-evolution of satellite galaxies and subhalos in observations will provide key clues to the picture of galaxy formation.

Measuring the masses of subhalos hosting satellite galaxies is a challenge, not only because dark matter does not emit light and can only be detected through its gravitational effects, such as gravitational lensing, but also because the subhalos hosting satellite galaxies have very small masses. In observations, the technique of strong gravitational lensing is employed to study the individual subhalos of lensing galaxies. These subhalos, distributed on the scale of the Einstein ring, can perturb the light path and manifest as flux-ratio anomalies \citep{Mao_and_Schneider1998, Metcalf_and_Madau2001, Nierenberg2014} or flux perturbations in the strong lensing images \citep[e.g.][]{Koopmans2005, Vegetti2009, Vegetti2010, Vegetti2012, LiRan2016, LiRan2017, HeQiuhan2022, HeQiuhan2023, Nightingale_arxiv220910566}. Such observations primarily involve dark matter halos with masses less than $10^{10} ~{\rm M_{\odot}}$. In the case of strong lensing by galaxy clusters, the dark matter halos of massive satellite galaxies can induce image displacements and variations in the brightness of extended arcs \citep[e.g.][]{Kneib1996, Kneib_Natarajan2011, Natarajan2009}. Although strong gravitational lensing can provide insights into the mass of individual subhalos, these events are rare and typically concentrated in the central regions of galaxies or galaxy clusters. Consequently, obtaining comprehensive measurements of the mass and evolution of satellite galaxy subhalos in galaxy groups and clusters remains challenging.

An alternative effective method for measuring the subhalos of satellite galaxies in galaxy groups and clusters is through the technique of galaxy-galaxy gravitational lensing, which measures tangential shear around a sample of selected galaxies\citep[e.g.][]{Brainerd_Blandford_Smail1996, Hoekstra_et_al_2003, Mandelbaum2005, Mandelbaum2006, LiRan2009, Cacciato2009, Mandelbaum_Seljak_Hirata_2008, Fuliping_Fanzuhui_RAA_review}. The measurement can probe the distribution of dark matter around the selected galaxy sample, thus helping to explore the connection between visible and invisible matter. In the context of galaxy-galaxy lensing, satellite galaxies can be selected from optically confirmed galaxy clusters or galaxy groups. By studying the gravitational lensing signal around these satellite galaxies, researchers can investigate the mass distribution of subhaloes, shedding light on the connection between the satellite galaxies and the subhalos in which they reside\citep[e.g.,][]{Yang2006_373_1159, Li2013}.

\citet{li2014} utilized data from the CFHT-STRIPE82 survey \citep[CS82][]{Comparat_2013MNRAS.433.1146C} and combined it with the SDSS galaxy group catalog constructed by \citet{Yang_2007ApJ...671..153Y}. They provided the first measurement of the galaxy-galaxy lensing signals for satellite galaxies. In \citet{li2016}, they further measured the lensing signals for satellite galaxies in the redMaPPer galaxy cluster catalog and found that the subhalo masses of satellite galaxies increase with their halo-centric radius, providing clear evidence of satellite galaxy mass loss. They also split the satellite galaxies into two mass bins and show that the satellite galaxies with larger stellar mass retain large dark matter subhalo. \citet{Sifón2015:1507.00737v3} measured the satellite galaxy lensing signals in the Galaxy And Mass Assembly survey \citep[GAMA;][]{Driver_2011MNRAS.413..971D} and found that while satellite galaxies exhibit significant mass loss compared to field galaxies, their stellar-to-halo-mass-ratio (SHMR) does not show a clear variation with halo-centric radius. \citet{Sifón2017:1706.06125v2} measured satellite galaxy-galaxy lensing with Multi-Epoch Nearby Cluster Survey \citep[MENeaCS][]{Sand_2012} and found a discontinuity trend of SHMR as a function of halo-centric radius. \citet{vanUitert2016} measured the galaxy-galaxy lensing signals in the GAMA survey and found no significant difference in the mass-to-light ratio between satellite galaxies and field galaxies. \citet{Niemiec2017:1703.03348v2} combined data from the CFHTLens survey, CS82 survey, and DES-SV survey to measure the gravitational lensing signals of satellite galaxies in the redMaPPer galaxy clusters. They confirmed that the mass-to-light ratio of satellite galaxies evolves with a halo-centric radius and calculated an average mass loss rate of approximately 70-80\% compared to field galaxies. Finally, \citet{Dvornik_2020A&A...642A..83D} measure the satellite galaxy-galaxy lensing for both central and satellite galaxies in the GAMA survey with shear catalog from  Kilo-Degree Survey, they confirmed that SHMR of satellite galaxies shifted toward lower halo masses by $\sim$20-50\% due to stripping mass loss. In summary, the results from different observational datasets show some discrepancies, indicating the need for improved data to accurately determine the evolution of subhalos hosting satellite galaxies in the environment of their host halos. 

In this project, we utilized the weak gravitational lensing measurements from the DECaLs survey \citep{Dey_2019}, covering an area of 9500 square degrees. We combined these measurements with the redMaPPer galaxy cluster catalog from the SDSS Data Release 8 \citep{Aihara2011} survey to perform galaxy-galaxy lensing measurements of satellite galaxies. This allowed us to obtain higher signal-to-noise ratio lensing signals for satellite galaxies,  calculate their subhalo mass, and derive their mass loss rates after infall more accurately.

The structure of our paper is as follows: In Section~\ref{sec:Observational_Data}, we introduce the observational data we used. In Section~\ref{sec:Methods}, we describe the methodology for galaxy-galaxy lensing calculations and lensing model. In Section~\ref{sec:Results_and_discussion}, we present our measurement results and discussion. Finally, in Section \ref{sec:summary_and_conclusions}, we provide our summary and conclusions. Throughout the paper, we adopt a flat $\Lambda$CDM cosmological model from the WMAP9 results \citep{Hinshaw_2013} (i.e., $\Omega_{\rm m}=0.2865$, $\rm H_{\rm 0}=69.32 \rm{kms^{-1}Mpc^{-1}}$).

\section{Observational Data}
\label{sec:Observational_Data}
%********************************
\begin{figure*}
\centering
\includegraphics[width=0.98\textwidth]{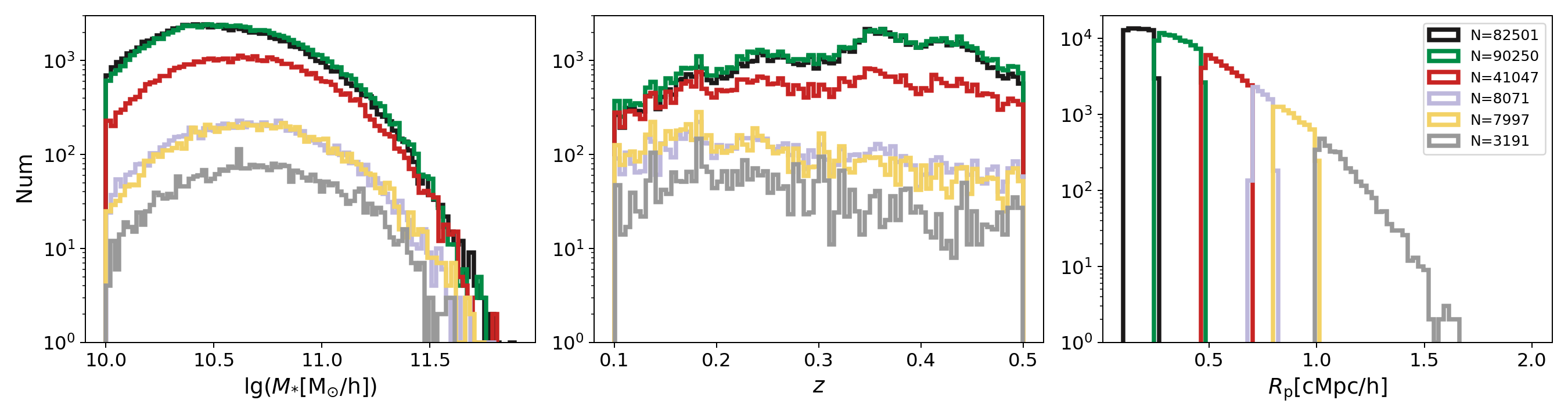}
\caption{Histogram of $M_{*}$, $z$, and $R_{\rm p}$ for the six bins listed in Table.~\ref{tab:table1}. The six bins  are shown in sequence from left to right in the third panel. Sub-samples in different panels share the same colors.} 
\label{fig:6bins}
\end{figure*}
%********************************

%*****************
\begin{figure*}
\centering
\includegraphics[width=0.49\textwidth]{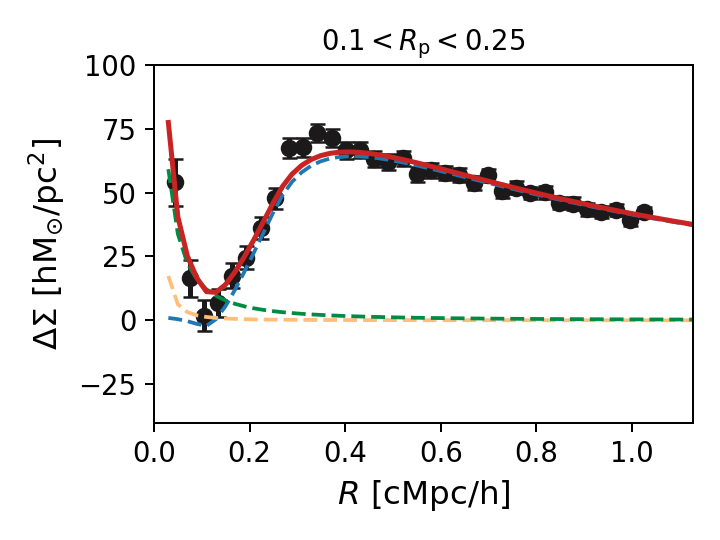}
\includegraphics[width=0.49\textwidth]{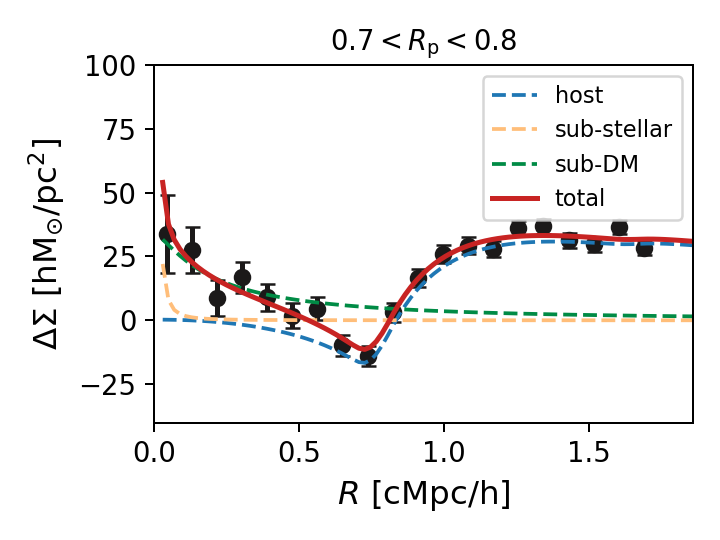}
\includegraphics[width=0.49\textwidth]{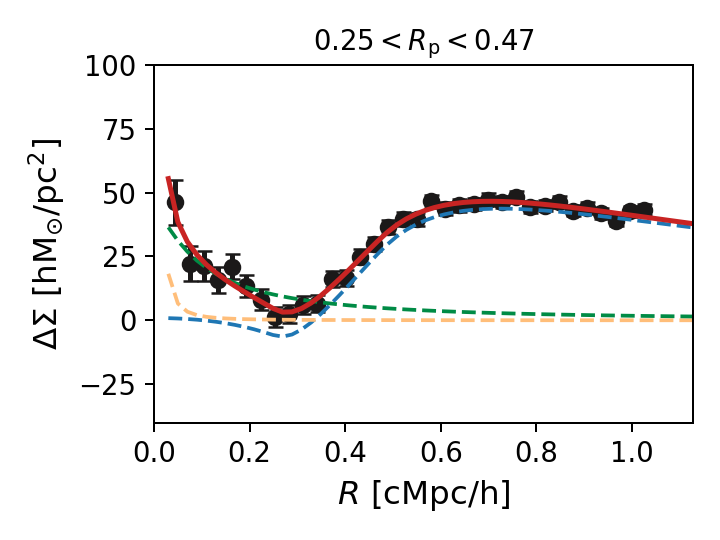}
\includegraphics[width=0.49\textwidth]{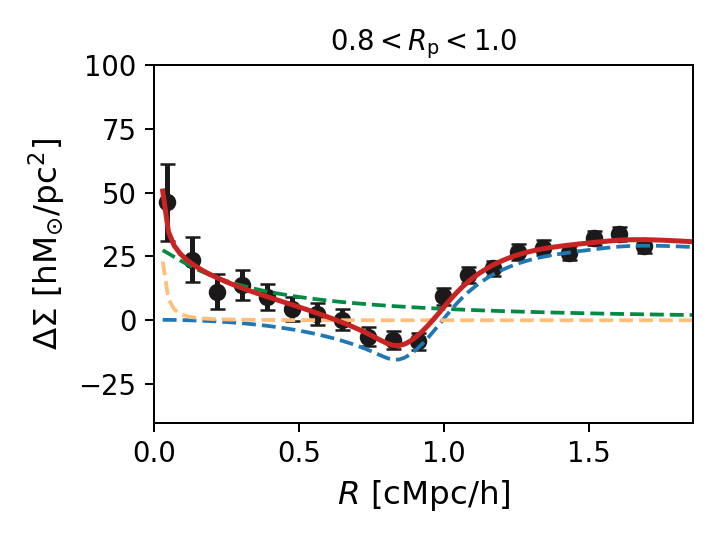}
\includegraphics[width=0.49\textwidth]{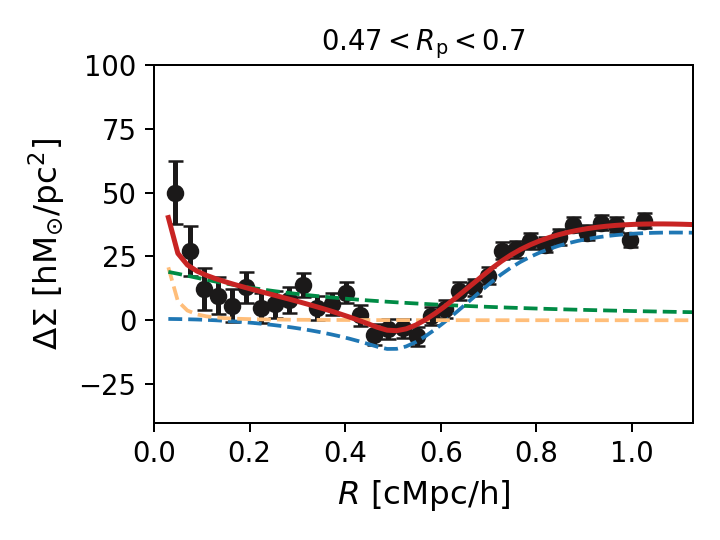}
\includegraphics[width=0.49\textwidth]{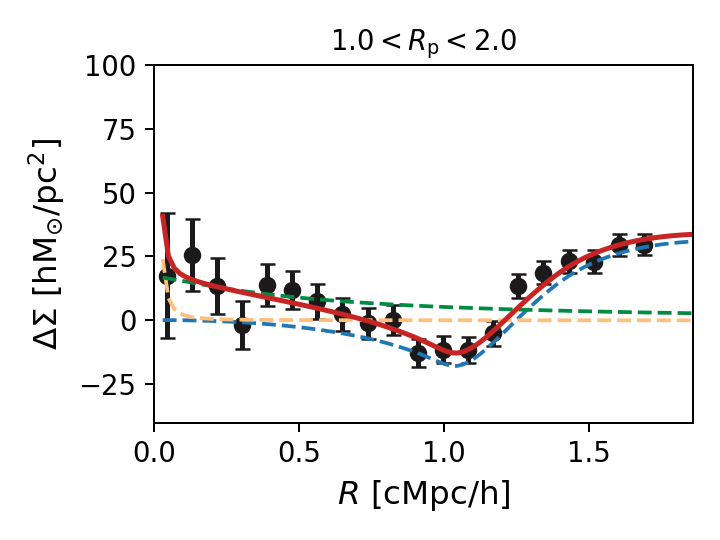}
\caption{This figure shows the stacked galaxy-galaxy subhalo lensing signal for each $R_{\rm p}$ bin and the corresponding best-fit model. The observed excess surface mass density $\Delta\Sigma(R)$ is represented by black circles with error bars, where the error bars reflect the 68 percent confidence intervals obtained using the jackknife resampling method. The best-fit model is shown as red lines, with the subhalo dark matter term represented by green lines, the stellar mass contribution from the satellite galaxy represented by orange lines, and the contribution from the host dark matter halo term represented by blue lines.}
\label{fig:Bestfit_Rp_photo}
\end{figure*}
%**************************

%************************
%
\begin{figure*}
\centering
\includegraphics[width=0.9\textwidth]{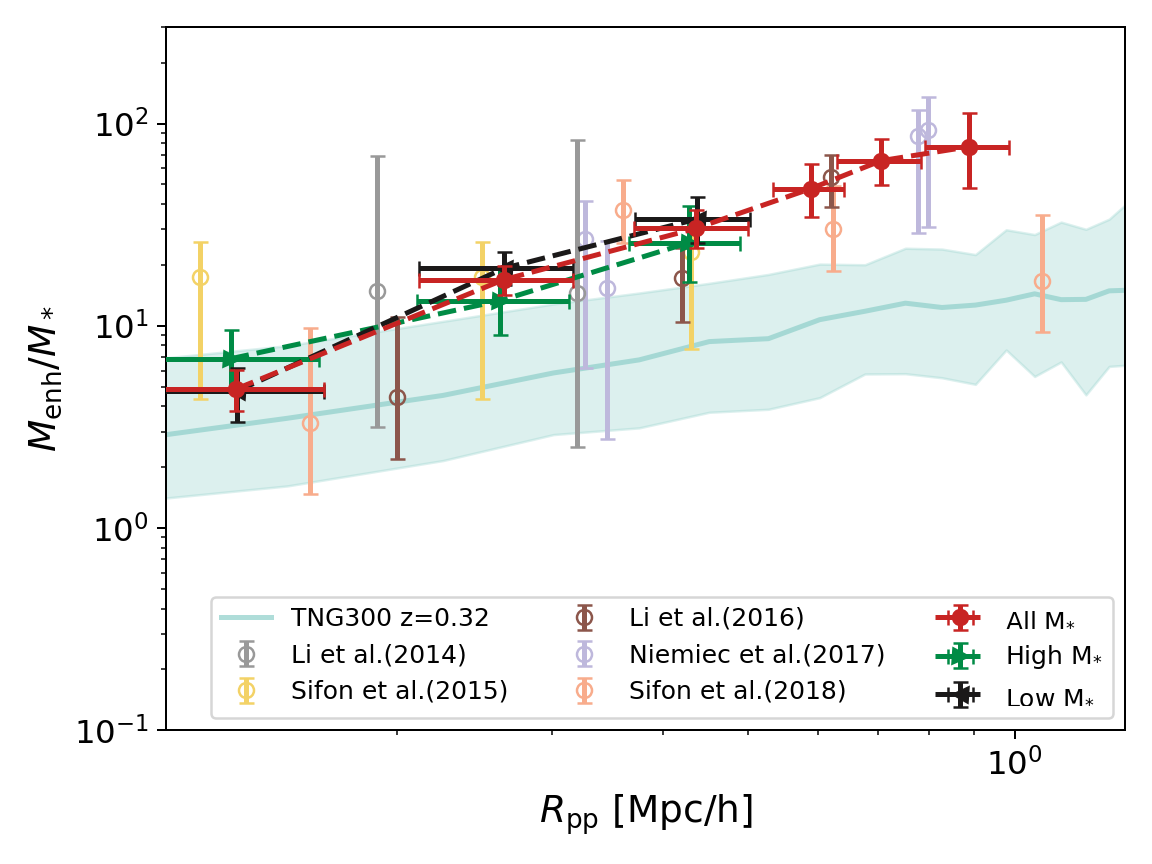}
\caption{This figure shows the evolution of SHMR of satellite galaxies with an increase of projected physical cluster-centric distance $R_{\rm pp}$. The red circles with error bars denote the best-fit SHMR measurement of this work. The green right triangle and black left triangle show the SHMR of our High-${\rm M}_{*}$ and Low-${\rm M}_{*}$ sub-samples. We compare our fitting result with the SHMR in TNG300 simulation of the IllustrisTNG project. The solid line represents the median and mean value of SHMR, and the upper and lower boundaries of the shaded area represent the 16th and 84th percentile. The other empty circles with error bars are the SHMR results from previous satellite galaxy-galaxy lensing observations\citep{li2014, Sifón2015:1507.00737v3, li2016, Niemiec2017:1703.03348v2, Sifón2017:1706.06125v2}.}
\label{fig:illustris300_shmr}
\end{figure*}
%**************************

%**************************
\begin{figure*}
\centering
\includegraphics[width=0.49\textwidth]{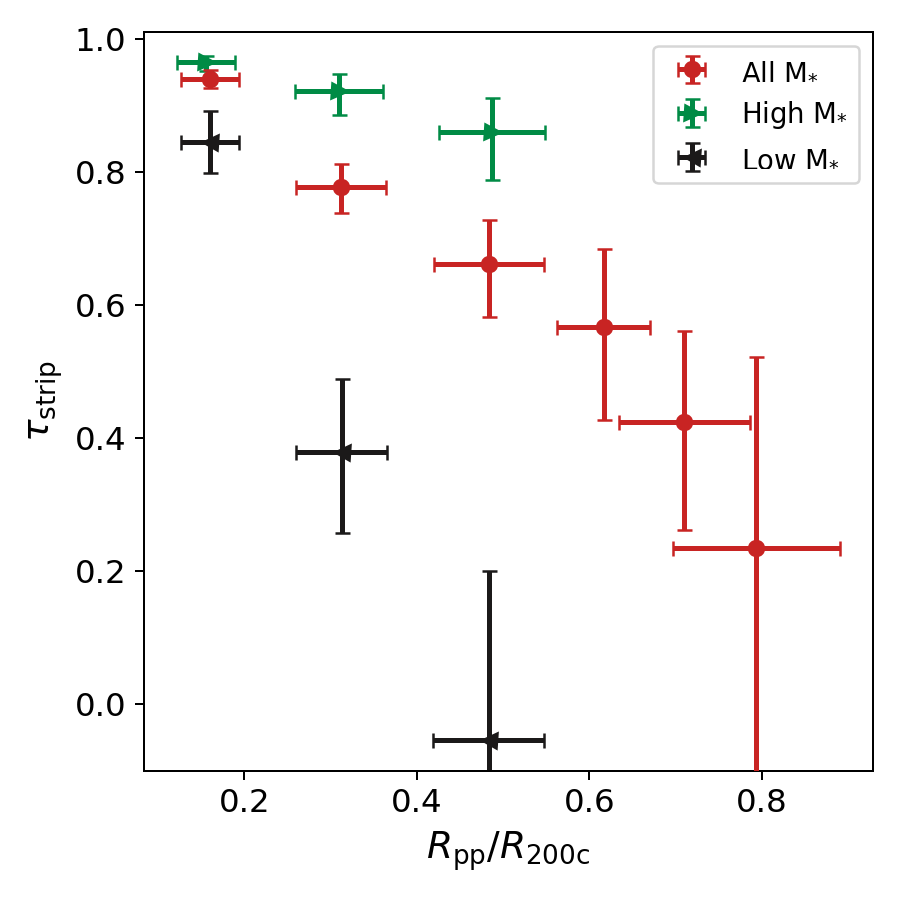}
\includegraphics[width=0.49\textwidth]{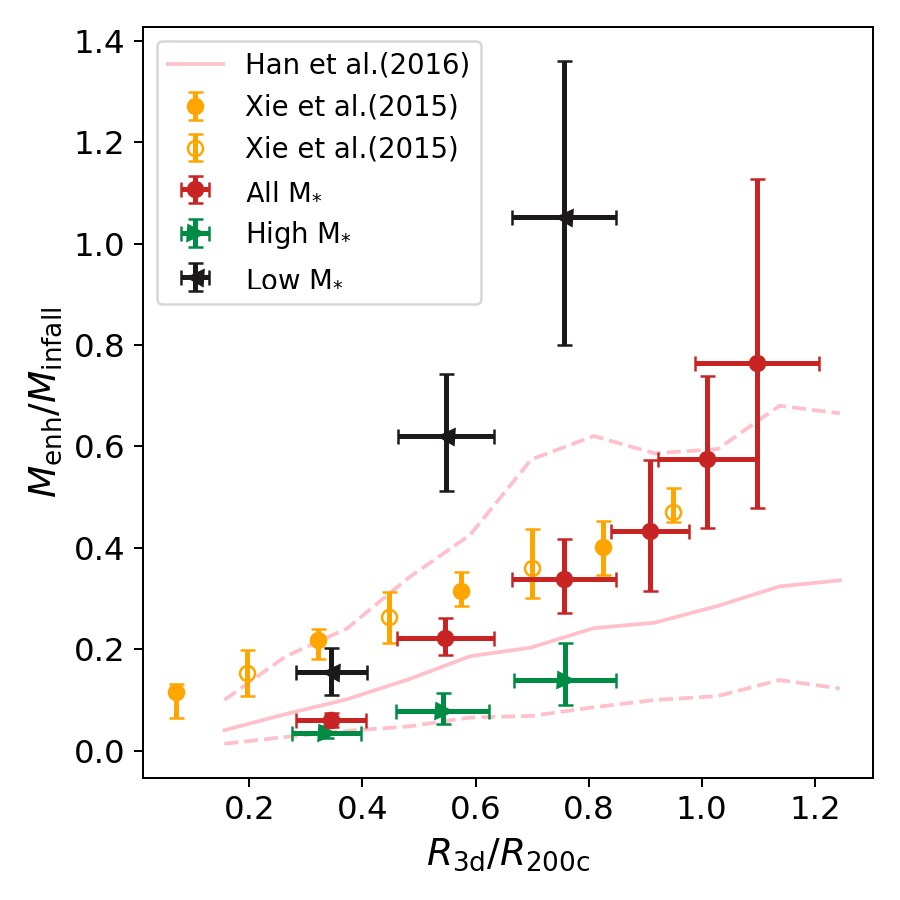}\\
\caption{Left: Mass loss rate of dark matter as a function of projected physical cluster-centric distance $R_{\rm pp}$. The red solid circles with error bars represent our results of sub-samples without binning by stellar mass, while the green triangles and black triangles represent the measurements for High-${\rm M}_{*}$ and Low-${\rm M}_{*}$, respectively. Right: The remained dark matter fraction as a function of three-dimensional cluster-centric distance $R_{\rm 3d}$ scaled with $R_{\rm 200c}$, with the same color scheme as in the left panel. The orange circles with error bars represent the Phoenix N-body simulation results taken from \citet{xielizhi_2015MNRAS.454.1697X}. The pink solid line and dashed line are from \citet{Han2016}, representing the median value of SHMR and $\pm 1\sigma$ confidence intervals, respectively.}
\label{fig:dm_loss_rate}
\end{figure*}
%**************************

%**************************
\begin{figure*}
\centering
\includegraphics[width=0.98\textwidth]{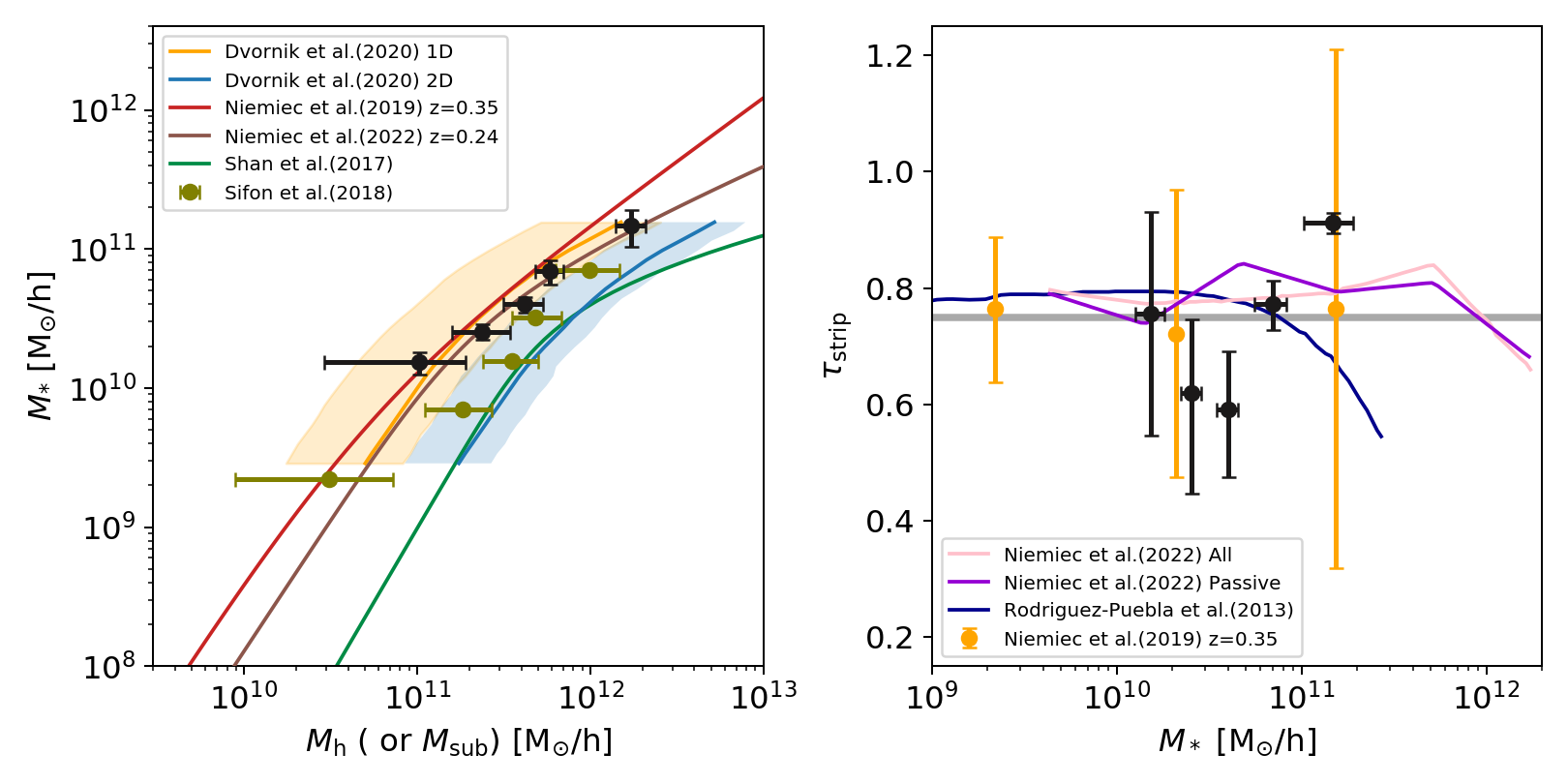}
\caption{Left panel: Relation between dark matter mass and stellar mass. The black solid circles with error bars represent the results of sub-samples binned by $M_{*}$ (see Appendix \ref{sec:a-b} for detailed sample binning). The red line represents the best-fit relation for dark matter mass and stellar mass of subhalos at $z=0.35$ in Illustris-1 \citep{Niemiec_2019MNRAS.487..653N}. The brown solid line represents the best-fit model for the stellar mass and dark matter mass of satellite galaxies at $z=0.24$ in TNG300 fitted by \citet{Niemiec2022MNRAS.512.6021N}. The green solid line represents the relation obtained by gravitational lensing measurements for the central/field galaxies in terms of their dark matter mass and stellar mass \citep{Shan_2017}. The orange (blue) solid line shows the relation between stellar mass and dark halo mass of satellite galaxies with weak gravitational lensing~\citep{Dvornik_2020A&A...642A..83D}. Right panel: Scatter plot of dark matter stripping rate versus stellar mass. The orange solid circles with error bars represent the average dark matter stripping rate of satellite galaxies with stellar masses between $2 \times 10^{7}~{\rm M_{\odot}/h}$ and $2 \times 10^{11}~{\rm M_{\odot}/h}$ in Illustris-1 at $z=0.35$. The grey horizontal line represents the average dark matter stripping rate of all satellite galaxies in Illustris-1 measured by \citep{Niemiec_2019MNRAS.487..653N}. The dark violet line shows the average dark matter stripping rate of passive satellite galaxies in TNG300 and the pink shows that of all satellite galaxies, both results come from \citet{Niemiec2022MNRAS.512.6021N}. The dark blue solid line represents the theoretical value of the dark matter stripping rate obtained by \citet{2013ApJ...767...92R}.}
\label{fig:averaged_striping_rate}
\end{figure*}
%**************************

In this project, we utilize satellite galaxies from the redMaPPer galaxy cluster as lenses and galaxies from the DECaLS Data Release 8 as sources. This section provides a description of these datasets.

\subsection{Lens galaxies}
\label{subsec:lens}
This study utilizes satellite galaxies in the redMaPPer cluster as gravitational lenses. The redMaPPer algorithm \citep[redMaPPer;][]{Rykoff2014, Rykoff2014a} groups red-sequence galaxies with similar redshifts and spatial concentrations based on their $ugriz$ magnitudes and errors to identify galaxy clusters.  In this work, we use version 6.3 of the redMaPPer cluster catalog\footnote{\url{http://risa.stanford.edu/redmapper/}} of SDSS Data Release 8 (DR8), which covers 10000 ${\rm deg}^2$ of the sky, contains 26,111 galaxy clusters \citep{Aihara2011}. In the redMaPPer catalog, each cluster is assigned a richness parameter $\lambda$ based on the number of red sequence galaxies brighter than $0.2L_{\rm *}$ at the cluster's redshift within a scaled aperture. This parameter has been shown to be a good proxy for the galaxy cluster halo mass \citep{Rykoff2014}. For this project, we select galaxy clusters with a richness $\lambda > 20$. We also require that our galaxy clusters reside within a redshift range of $0.1 < z < 0.5$, where the lower bound ensures lensing efficiency and the higher bound ensures reliable richness measurements \citep{Rozo2014}.

For each redMaPPer cluster, the potential member is  assigned a probability of membership $P_{\rm mem}$ according to their photometric redshift, color, and their cluster-centric distance. To reduce the contamination induced by fake member galaxies, we only use satellite galaxies with membership probability $P_{\rm mem} > 0.8$ and this selection criterion can remove most contamination \citep{Zu_2017_10.1093/mnras/stx1264, Niemiec2017:1703.03348v2}. 

When calculating the lensing signal, we use the redshift of the central galaxy of each redMaPPer galaxy cluster as the redshift of the satellite galaxies, as the majority of central galaxies have spectroscopic redshifts. We make use of the stellar mass information  derived by \citet{Zou2019}, where the stellar mass is estimated by applying the Bayesian spectral energy distribution (SED) model fitting with the Le Phare code\footnote{\url{http://www.cfht.hawaii.edu/~arnouts/LEPHARE/lephare.html}} \citep{Ilbert2009}. \citet{Zou2019} adopted the default BC03 spectral models with the \citet{Chabrier_2004} IMF. Readers are referred to \cite{Zou2019} for more details. In this project, we select satellite galaxies within a stellar mass region of [$10^{10}~{\rm M_{\odot}/h}$,$10^{12}~{\rm M_{\odot}/h}$].

We divide the satellite galaxies into six bins according to their comoving projection cluster-centric distance $R_{\rm p}$. The ranges of $R_{\rm p}$ bins and the number of satellite galaxy lenses in each bin are shown in Table.~\ref{tab:table1}. We show the distribution of stellar mass $M_{*}$, redshift $z$, and comoving projected cluster-centric distance $R_{\rm p}$ of each bin in Fig.~\ref{fig:6bins}.

\subsection{Source galaxies}
\label{ssec:DECaLS_survey}
    The source galaxies catalog for weak lensing analysis is extracted from data release 8 (DR8) of the Dark Energy Spectroscopic Instrument (DESI) Legacy Imaging Surveys \citep[DECals,][]{Dey_2019}, and  has been used in multiple scientific studies \citep[e.g.][]{Phriksee2020, Yao2020, Zu2021,xuweiwei_2021,wang_2023arXiv230411715W}, due to its large sky coverage of approximately 9500 ${\rm deg}^2$ in $grz$ bands.

    The DECaLS DR8 data is processed by {\it Tractor}~\citep{Meisner2017,Lang2016AJ....151...36L}. The morphologies of sources are divided into five types, including point sources (PSF), simple galaxies (SIMP, an exponential profile with affixed $0\farcs45$ effective radius and round profile), DeVaucouleurs (DEV, elliptical galaxies), Exponential (EXP, spiral galaxies), and Composite model (COMP, deVaucouleurs + exponential profile with the same source center)\footnote{\url{https://www.legacysurvey.org/dr8/description/}}. Sky-subtracted images are stacked in five different ways: one stack per band, one flat Spectral Energy Distribution (SED) stack of the $g$, $r$, $z$ bands,  and one red SED stack of all bands ($g-r=1$~mag and $r-z=1$~mag). Sources above the 6$\sigma$ detection limit in any stack are kept as candidates. Galaxy ellipticities (e1,e2) are estimated by a joint fitting image of $g$, $r$, and $z$ bands for  SIMP, DEV, EXP, and COMP galaxies. The multiplicative bias ($m$) and additive biases \citep[e.g.][]{Heymans2012MNRAS.427..146H, Miller2013} are modeled by calibrating with the image simulation \citep{Phriksee2020} and cross-matching  with external shear measurements \citep{Phriksee2020,Yao2020,Zu2021}, including the Canada-France-Hawaii Telescope (CFHT) Stripe 82 \citep{Moraes2014}, Dark Energy Survey \citep{DES2016}, and Kilo-Degree Survey \citep{Hildebrandt2017} objects.

	The photo-$z$ of each source galaxy in DECaLS DR8 shear catalog is taken from  \citet{Zou2019}, where the redshift of a target galaxy is derived with its k-nearest-neighbor in the SED space whose spectroscopic redshift is known. The photo-z is derived using 5 photometric bands: three optical bands, $g$, $r$, and $z$ from DECaLS DR8, and two infrared bands, W1, W2, from Wide-Field Infrared Survey Explorer (WISE). By comparing with a spectroscopic sample of 2.2 million galaxies, \citet{Zou2019} shows that the final photo-z catalog has a redshift bias of $\Delta \overline{z}_{\rm norm}=2.4\times10^{-4}$, the accuracy of $\sigma_{\Delta z_{\rm norm}}=0.017$, and outlier rate of about 5.1\%.

%%%%%%%%%%%%%%%%%%%%%%%%%%%%%   Methods   %%%%%%%%%%%%%%%%%%%%%%%%%%%%
\section{Methods}
\label{sec:Methods}
\subsection{Lensing signal}
The excess surface density, $\Delta \Sigma (R)$ is calculated as
\begin{equation}
\Delta \Sigma (R) =\overline{\Sigma}(<R)-\overline{\Sigma}(R)=\frac{\sum_{\rm ls}\omega_{\rm ls}\gamma_{\rm t}^{\rm ls}\Sigma_{\rm crit}}{\sum_{\rm ls}\omega_{\rm ls}}\,,
\end{equation}
\noindent where
\begin{equation}
\omega_{\rm ls}=\omega_{\rm n}\Sigma_{\rm crit}^{-2}\,,
\end{equation}

\begin{equation}
\Sigma_{\rm crit}=\frac{c^2}{4 \rm \pi \rm G}\frac{D_{\rm s}}{D_{\rm l}D_{\rm ls}}\,.
\end{equation}

\noindent $\overline{\Sigma}(<R)$ is the mean density within radius $R$ and the $\overline{\Sigma}(R)$ is the azimuthally averaged surface density at radius $R$~\citep[e.g.][]{Miralda-Escude1991ApJ...370....1M,Wilson2001ApJ...555..572W,Leauthaud2010ApJ...709...97L}. Here, $\gamma_{\rm t}$ is the tangential shear, and $\Sigma_{\rm crit}$ is the critical surface density containing space geometry information. Here, $D_{\rm s}$, $D_{\rm l}$, and $D_{\rm ls}$ are the angular diameter distances between the observer and the source, the observer and the lens, and the source and lens, respectively. The $c$ here is the constant of
light velocity in the vacuum.
$\omega_{\rm n}$ is a weight factor introduced to account for intrinsic scatter in ellipticity and shape measurement error of each source galaxy \citep{Miller2007, Miller2013}. The $\omega_{\rm n}$ we used in this work is defined as $\omega_{\rm n}=1/(\sigma^2_{\epsilon}+\sigma^2_{\rm e})$.  $\sigma_{\epsilon}=0.27$ is the intrinsic ellipticity dispersion derived from the whole galaxy catalog \citep{Giblin2021}. $\sigma_{\rm e}$ is the error of the ellipticity measurement defined in \citet{Hoekstra2002}. Owing to the  photo-$z$ uncertainties of the source galaxies, we remove the lens-source pairs with $z_{\rm s}-z_{\rm l}< 0.1$ or $z_{\rm s}-z_{\rm l}<\sigma_{\rm l} + \sigma_{\rm s}$. $\sigma_{\rm l}$ and $\sigma_{\rm s}$ are redshfit errors of lens and source, respectively.

We apply the correction of multiplicative bias to the measured excess surface density as
\begin{equation}
\Delta \Sigma^{\rm cal}(R)=\frac{\Delta \Sigma (R)}{1+K(z_{\rm l})} \,,
\end{equation}
where
\begin{equation}
1+K(z_{\rm l})=\frac{\sum_{\rm ls}\omega_{\rm ls}(1+m)}{\sum_{\rm ls}\omega_{\rm ls}}\,.
\end{equation}
where m is the multiplicative error as described in Sec.~\ref{ssec:DECaLS_survey}. 
In this work, we use the Super W Of Theta (\textsc{SWOT}) code\footnote{\url{http://jeancoupon.com/swot}}~\citep{Coupon_2011} to calculate the excess surface density.

We stack the tangential shear around satellite galaxies in 6 subsamples of $R_{\rm p}$ bins as listed in Table.~\ref{tab:table1}. For subsamples of $0.1<R_{\rm p}<0.25$, $0.25<R_{\rm p}<0.47$  and $0.47<R_{\rm p}<0.7$, we calculate galaxy-galaxy in 35 linear radial bins ranging from 0.05 to 1 Mpc/h in comoving coordinates. For the larger $R_{\rm p}$ bins, we use 20 linear radial bins ranging from 0.05 to 1.75 Mpc/h in comoving coordinates.

\subsection{Lensing model}
\label{subsec:lensing_model}
The excess surface density around a satellite galaxy is composed of three components:
\begin{eqnarray}
\Delta \Sigma (R)=\Delta \Sigma _{\rm sub}(R)+\Delta \Sigma _{\rm host}(R, R_{\rm p})+\Delta \Sigma _{\rm star}(R)\,,
\end{eqnarray}
where the $\Delta \Sigma _{\rm sub}$ is the contribution from the subhalo in which the satellite galaxy resides, $\Delta \Sigma _{\rm host}$ is the contribution from the host halo of the cluster, where $R_{\rm p}$ is the projected distance from the satellite galaxy to the center of the host halo, and $\Delta \Sigma _{\rm star}$ is the contribution from the stellar component of the satellite galaxy.
Since the contribution from the two-halo term is only significant at $R>3~{\rm Mpc/h}$ for clusters \citep{Shan_2017}, it cannot affect the region where satellite galaxies dominate. Therefore, we have neglected the two-halo term.

\begin{itemize}
\item{Subhalo contribution}
\end{itemize}
Different mass density models of subhalo were studied using gravitational lensing \citep{li2016, Sifón2015:1507.00737v3, Sifón2017:1706.06125v2, Niemiec2017:1703.03348v2}. The two most commonly used models are the NFW model \citep{Navarro_1997} and the truncated-NFW (tNFW) profile \citep{Edward_A.Baltz_2009,Oguri_2011MNRAS.414.1851O}. In this study, we choose the NFW profile model as the subhalo mass density model.
\begin{equation}
\rho (r) = \frac{\rho_{\rm crit}\delta_{\rm crit}}{(r/r_{\rm s})(1+r/r_{\rm s})^2}\,,
\end{equation}
where $r_{\rm s}$ is the characteristic scale of the halo  where the local logarithmic slope reaches $\frac{{\rm d}~\ln\rho}{{\rm d}~\ln r}=-2$. The critical density of the universe is written as
\begin{equation}
  \rho_{\rm crit}=\frac{3H(z)^2}{8\pi G}\,, 
\end{equation}
where $H(z)$ is Hubble parameter at redshift $z$ and the $G$ is Newton’s constant. 
\begin{equation}
    \delta_{\rm crit}=\frac{\Delta}{3}\frac{C_{\Delta}^3}{{\rm ln}(1+C_{\Delta})-C_{\Delta}/(1+C_{\Delta})}\,,
\end{equation}
$C_{\Delta}=R_{\Delta}/r_{\rm s}$
is the concentration parameter, $R_{\Delta}$ is a radius where the average density of the halo within it is $\Delta$ times of the mean matter mass density  $\rho_{\rm crit}\Omega_{\rm m}(z)$ of the universe at redshift $z$, where $\Omega_{\rm m}(z)$ is the matter density  parameter at redshift $z$. The enclosed mass within $R_{\Delta}$ is $M_{\Delta}=\frac{4\Delta\pi}{3}\rho_{\rm crit}\Omega_{\rm m}(z)R_{\Delta}$. In this study, we choose $\Delta=200$. The free parameters of this model are $M_{\rm 200m}$ and $C_{\rm 200m}$. The corresponding halo radius is $R_{\rm 200m}$. In the latter part of the paper, we also use another definition of halo radius $R_{\rm 200c}$, which represents the radius within which the mean density of the halo is 200 times the critical density of the universe at the redshift $z$ the halo located. The corresponding mass and concentration are denoted as $M_{\rm 200c}$ and $C_{\rm 200c}$.

By integrating the three-dimensional (3D) density profile along the line of sight, we can get the projected surface density $\Sigma_{\rm NFW}(R)$ which is a function of the projection radius $R$,
\begin{equation}
\Sigma _{\rm NFW}(R)=\int _{-\infty}^{\infty } \rho \left(\sqrt{R^2+z^2}\right)\text{d}z\,.
\end{equation}

Integrating $\Sigma_{\rm NFW}$(R) from 0 to $R$, we can get the mean surface density within $R$, $\overline{\Sigma} _{\rm NFW}(<R)$, as follow,

\begin{equation}
\overline{\Sigma} _{\rm NFW}(<R)=\frac{2}{R^2}\int _{0}^{R}R^{\prime }\Sigma _{\rm NFW}(R^{\prime })\,\text{d}R^{\prime}\,,
\end{equation}
The lensing signal produced by the NFW profile is 
\begin{equation}
   \Delta\Sigma(R)=\overline{\Sigma}_{\rm NFW}(<R)-\Sigma_{\rm NFW}(R)\,.  
\end{equation}

Note that the quantity $M_{\rm 200m}$ and $C_{\rm 200m}$ of the subhalo density profile are used for mathematical convenience only, not physically meaningful for subhaloes whose outer part has been stripped in their host haloes. In this paper, we define subhalo masses, $M_{\rm enh}$ as the sum of dark matter mass within the subhalo radius, $r_{\rm sub}$, at which the subhalo dark matter mass density equals to the background mass density of the cluster \citep{Natarajan2007MNRAS.376..180N, Sifón2017:1706.06125v2}. The subhalo radius $r_{\rm sub}$ is determined by measuring the mean mass density within a small sphere around the substructure and subtracting from it the mass in the same sphere after spherically averaging the entire mass distribution of the halo around the halo center. This provides an estimate of the background density in the volume occupied by the substructure. During the computation of $r_{\rm sub}$, it is necessary to have knowledge of the three-dimensional halo-centric radius $R_{\rm 3d}$. Assuming that the satellite galaxy number density distribution follows the NFW model distribution and is consistent with the distribution of dark matter particles in the host halo, then statistically, the average of the three-dimensional cluster-centric distance of the dark matter particles (satellite galaxies) projected onto the $R_{\rm p}$ radius can be expressed as follow,
\begin{equation}
        \overline{R}_{\rm 3d}=\frac{\int_{-a}^{+a} r \rho(r)\text{d} z}{\int_{-a}^{+a}  \rho(r)\text{d} z}\,,   
\end{equation}
where $a=\sqrt{(3R_{\rm 200m,host})^2-R_{\rm p}^2}$ and $\rho(r)$ is the mass density profile of host halo. The mass and concentration of the host halo mass model are shown in the following host halo model part.

\begin{itemize}
\item{Host halo model}
\end{itemize}

We assume that the profile of a host halo in a galaxy cluster follows the NFW profile, the contribution from the host halo can be expressed as follows according to \cite{Yang2006_373_1159}.

\begin{gather}
\Delta\Sigma_{\rm NFW,host}=\overline{\Sigma} _{\rm NFW, host}\left(<R|R_{\rm p}\right)-\Sigma _{\rm NFW, host}\left(R|R_{\rm p}\right)\,, \notag \\
\Sigma _{\rm NFW, host}(R|R_{\rm p})=\ \int _{0}^{2\pi } \text{d}\theta \Sigma _{\rm NFW, host}\left(\sqrt{R^{2}+R_{\rm p}^{2}+2RR_{\rm p}{\rm cos}(\theta )}\right)\,, \notag \\
\overline{\Sigma} _{\rm NFW,host}\left(<R|R_{\rm p}\right)=\frac{2}{R^2}\int _{0}^{R}R^{\prime }\Sigma _{\rm NFW, host}\left(R^{\prime }|R_{\rm p}\right)\,\text{d}R^{\prime}\,. 
\end{gather}

To calculate the lensing signal for each galaxy cluster, the values of $M_{\rm 200m}$ and $C_{\rm 200m}$ of host halo are obtained through the $\lambda-M_{\rm 200m}$ relation presented by \citet{Rykoff2012}, 

\begin{equation}
{\rm ln} \left ( \frac{M_{\rm 200m}}{h_{70}^{-1}10^{14}M_{\odot}}\right )=1.72+1.08{\rm ln}\left(\lambda/60\right)\,,
\end{equation}
as well as the $M_{\rm 200m}-C_{\rm 200m}$ relation proposed by \cite{xuweiwei_2021}\,,
\begin{equation}
C_{\rm 200m}=C_{0}\left( \frac{M_{\rm 200m}}{10^{12}M_{\odot}/h}\right )^{-\gamma}\Bigg [ 1+\left(\frac{M_{\rm 200m}}{M_{0}}\right)^{0.4}\Bigg ]\,,
\end{equation}
 \noindent  where $C_{0}=5.119^{+0.183}_{-0.185}$, $\gamma=0.205^{+0.010}_{-0.010}$, ${\rm lg(}M_{0}{\rm )}=14.083^{+0.130}_{-0.133}$  when $0.08<z<0.35$ and $C_{0}=4.875_{-0.208}^{+0.209}$, $\gamma=0.221_{-0.010}^{+0.010}$,  ${\rm lg(}M_{0}{\rm )}=13.750^{+0.142}_{-0.141}$  when $0.35<z<0.65$.
 In the redMaPPer catalog, each cluster has five possible central
galaxies, each with probability $P_{\rm cen}$. For each probable satellite-central galaxy pair, we calculate $\Delta\Sigma_{\rm host, i, j}$. Then we get the average contribution of host halo in each sub-samples as 
\begin{equation}
\Delta\Sigma_{\rm host}=\frac{\alpha}{N_{\rm sat}}\sum _{i}^{N_{\rm sat}}\sum_{j}^{5}\Delta\Sigma_{\rm host,i,j}\left(R|R_{\rm p,i,j}\right)P_{\rm cen,i,j}\,,
\end{equation}
where $R_{\rm p, i, j}$ is the projection distance between the i-th satellite galaxy and its j-th host galaxy cluster center, and the $P_{\rm cen, i, j}$ is the corresponding  probability of the central galaxy being the central galaxy. $\alpha$ is the only free parameter in the host halo model that can adjust the lensing amplitude. If the richness-mass and mass-concentration  relations are perfect, the best-fit of $\alpha$ should be close to unity.

\begin{itemize}
\item{Satellite stellar contribution}
\end{itemize}

The lensing contributed from the stellar component within subhalos is usually modelled as a point mass:  
\begin{equation}
\Delta \Sigma _{\rm star}(R)=\frac{ \left<M_{\rm *}\right>}{\pi R^2}\,,
\end{equation}
where the $M_{\rm *}$ is the stellar mass in subhalos. Here we use the average stellar mass of stacked satellite galaxies lens $\left< M_{*} \right>$.

We fit our model to the observational data with three free parameters $\alpha$, $M_{\rm 200m}$, and $C_{\rm 200m}$ in the model.

\section{Results and discussion}
\label{sec:Results_and_discussion}
We use the Markov Chain Monte Carlo sampler \textsc{Emcee} \footnote{\url{https://emcee.readthedocs.io/en/stable/}}\citep{2013PASP..125..306F} to fit the weak lensing signal  to get the posterior distribution of the free parameters. We use 120 chains of 300,000 steps. A uniform distribution is adopted for each free parameter:
\begin{itemize}
\item{${\rm 10^7~{\rm M_{\odot}/h}}<M_{\rm 200m}<10^{14}~{\rm M_{\odot}/h}$}
\item{${\rm 0< }C_{\rm 200m}{\rm < 40}$}
\item{${0<\alpha<2}$}
\end{itemize}

We present the galaxy-galaxy lensing signal of satellite galaxies in different $R_{\rm p}$ bins, along with their corresponding best-fit models in Fig.~\ref{fig:Bestfit_Rp_photo}. The excess surface mass density $\Delta\Sigma(R)$ of the cluster sample is represented by black circles with error bars, where the error bars reflect the 68 percent confidence intervals obtained using jackknife resampling. The best-fit models are shown as red solid lines, and the different components of the best-fit model are represented by orange (stellar component), green (subhalo dark matter), and blue (host halo) lines, respectively. The model fitting results are listed in Table.~\ref{tab:table1}. The fitted value of the host halo normalization parameter $\alpha$ is very close to 1, indicating that the host-halo contribution is very well described.

\begin{table*}
\begin{center}
\caption{Number of lenses in each sub-sample. The bins are separated with the $R_{\rm p}$ value, as shown in the second column. In the following columns, the parameters of each bin are listed in sequence, the number of lenses of sub-sample, the average redshift, the average projection cluster-centric physical distance, the average comoving projection cluster-centric physical distance, the average stellar mass, host halo normalization $\alpha$, subhalo mass, SHMR and dark matter strip rate. All the masses are in unit of ${\rm M_{\odot}/h}$ and distance in Mpc/h. }
\begin{tabular}{llllllllllll}
\hline
\hline
 & $R_{\rm p}$ range & Num  & $\left < z \right >$ & $\left < R_{\rm pp}\right>$  &   $\left < R_{\rm p} \right >$  &  ${\rm lg} \left ( \left <M_{*} \right > \right)$  & $\alpha$  & $r_{\rm sub}$  & ${\rm lg} \left ( M_{\rm enh} \right)$  & $M_{\rm enh}/\left < M_{*} \right>$   & $\tau_{\rm strip}$ \\
\hline
${\rm All~M_{*}}$ &0.1-0.25  &  82501  &  0.33  &  0.13  &  0.17 & 10.69  & $0.99_{-0.01}^{+0.01}$  &$0.05_{-0.01}^{+0.01}$  &    $11.38_{-0.11}^{+0.09}$  & $4.87_{-1.08}^{+1.15}$  &  $0.94_{-0.01}^{+0.01}$ \\
  &  0.25-0.47  &  90250  &  0.33  &  0.26  &  0.35 & 10.71  & $0.98_{-0.02}^{+0.02}$  &$0.14_{-0.01}^{+0.01}$  &    $11.94_{-0.07}^{+0.07}$  & $16.89_{-2.63}^{+2.94}$  &  $0.78_{-0.04}^{+0.03}$ \\
  &  0.47-0.7  &  41047  &  0.31  &  0.44  &  0.57 & 10.77  & $0.99_{-0.04}^{+0.04}$  &$0.25_{-0.03}^{+0.03}$  &    $12.25_{-0.1}^{+0.09}$  & $30.27_{-5.97}^{+7.11}$  &  $0.66_{-0.08}^{+0.07}$ \\
  &  0.7-0.8  &  8071  &  0.28  &  0.59  &  0.75 & 10.79  & $1.03_{-0.04}^{+0.04}$  &$0.3_{-0.04}^{+0.04}$  &    $12.47_{-0.14}^{+0.12}$  & $47.58_{-12.96}^{+15.3}$  &  $0.57_{-0.14}^{+0.12}$ \\
  &  0.8-1.0  &  7997  &  0.26  &  0.71  &  0.88 & 10.81  & $0.98_{-0.05}^{+0.04}$  &$0.38_{-0.05}^{+0.06}$  &    $12.63_{-0.12}^{+0.11}$  & $65.51_{-15.59}^{+18.43}$  &  $0.42_{-0.16}^{+0.14}$ \\
  &  1.0-2.0  &  3191  &  0.26  &  0.89  &  1.12 & 10.83  & $1.03_{-0.09}^{+0.08}$  &$0.47_{-0.09}^{+0.11}$  &    $12.72_{-0.2}^{+0.17}$  & $76.78_{-28.75}^{+36.54}$  &  $0.24_{-0.36}^{+0.29}$ \\
\hline
${\rm High~M_{*}}$ &0.1-0.25  &  8170  &  0.35  &  0.13  &  0.17 & 11.18  & $0.98_{-0.05}^{+0.04}$  &$0.09_{-0.01}^{+0.02}$  &    $12.02_{-0.15}^{+0.14}$  & $6.87_{-2.0}^{+2.67}$ &  $0.96_{-0.01}^{+0.01}$ \\
  &  0.25-0.47  &  9943  &  0.35  &  0.26  &  0.35 & 11.18  & $0.95_{-0.07}^{+0.06}$  &$0.18_{-0.03}^{+0.04}$  &    $12.3_{-0.17}^{+0.16}$  & $13.31_{-4.32}^{+6.04}$ &  $0.92_{-0.04}^{+0.03}$ \\
  &  0.47-0.7  &  5955  &  0.34  &  0.43  &  0.57 & 11.18  & $1.11_{-0.12}^{+0.11}$  &$0.27_{-0.07}^{+0.08}$  &    $12.6_{-0.19}^{+0.18}$  & $25.74_{-9.3}^{+13.21}$ &  $0.86_{-0.07}^{+0.05}$ \\
\hline
${\rm Low~M_{*}}$ &0.1-0.25  &  8170  &  0.33  &  0.13  &  0.17 & 10.58  & $0.99_{-0.01}^{+0.01}$  &$0.05_{-0.01}^{+0.01}$  &    $11.25_{-0.15}^{+0.12}$  & $4.74_{-1.39}^{+1.46}$ &  $0.85_{-0.05}^{+0.05}$ \\
  &  0.25-0.47  &  9943  &  0.33  &  0.26  &  0.35 & 10.59  & $0.99_{-0.02}^{+0.02}$  &$0.13_{-0.01}^{+0.01}$  &    $11.88_{-0.08}^{+0.08}$  & $19.24_{-3.38}^{+3.79}$ &  $0.38_{-0.12}^{+0.11}$ \\
  &  0.47-0.7  &  5955  &  0.3  &  0.44  &  0.56 & 10.63  & $0.98_{-0.05}^{+0.04}$  &$0.24_{-0.03}^{+0.04}$  &    $12.16_{-0.12}^{+0.11}$  & $33.67_{-8.11}^{+9.81}$ &  $-0.05_{-0.31}^{+0.25}$ \\
\hline
\end{tabular}
\label{tab:table1}
\end{center}
\end{table*}

\begin{table*}
\begin{center}
\caption{Number of lenses in each sub-sample. The bins are separated with the $M_{*}$, as shown in the first column. In the following columns, the parameters of each bin are listed in sequence, the number of lenses of sub-sample, the average redshift, the average projection cluster-centric physical distance, the average comoving projection cluster-centric physical distance, the average stellar mass, host halo normalization $\alpha$, subhalo mass, SHMR and average dark matter strip rate. All the masses are in unit of ${\rm M_{\odot}/h}$ and distance in Mpc/h.}
\begin{tabular}{lllllllllll}
\hline
\hline
${\rm lg(}{M_{*}}{\rm )}$ range & Num  & $\left < z \right >$ & $\left < R_{\rm pp}\right>$  &   $\left < R_{\rm p} \right >$  &  ${\rm lg} \left ( \left <M_{*} \right > \right)$  & $\alpha$  & $r_{\rm sub}$  & ${\rm lg} \left ( M_{\rm enh} \right)$  & $M_{\rm enh}/\left < M_{*} \right>$  & $\tau_{\rm strip}$ \\
\hline
10.0-10.3  &  42186  &  0.3  &  0.25  &  0.33 & 10.19  & $1.04_{-0.02}^{+0.01}$  &$0.05_{-0.02}^{+0.02}$  &    $11.01_{-0.55}^{+0.27}$  & $6.71_{-4.8}^{+5.78}$&  $0.76_{-0.21}^{+0.17}$ \\
10.3-10.5  &  51329  &  0.31  &  0.26  &  0.34 & 10.41  & $1.01_{-0.03}^{+0.01}$  &$0.07_{-0.01}^{+0.02}$  &    $11.38_{-0.18}^{+0.16}$  & $9.37_{-3.13}^{+4.23}$&  $0.62_{-0.17}^{+0.13}$ \\
10.5-10.7  &  51736  &  0.32  &  0.27  &  0.36 & 10.6  & $0.98_{-0.02}^{+0.02}$  &$0.09_{-0.01}^{+0.01}$  &    $11.62_{-0.12}^{+0.11}$  & $10.41_{-2.53}^{+3.0}$&  $0.59_{-0.12}^{+0.1}$ \\
10.7-11.0  &  57828  &  0.33  &  0.29  &  0.38 & 10.84  & $0.98_{-0.01}^{+0.01}$  &$0.09_{-0.01}^{+0.01}$  &    $11.76_{-0.08}^{+0.08}$  & $8.36_{-1.45}^{+1.66}$&  $0.77_{-0.05}^{+0.04}$ \\
11.0-11.5  &  26241  &  0.34  &  0.3  &  0.39 & 11.17  & $0.96_{-0.03}^{+0.03}$  &$0.15_{-0.02}^{+0.02}$  &    $12.24_{-0.09}^{+0.09}$  & $11.67_{-2.16}^{+2.55}$&  $0.91_{-0.02}^{+0.02}$ \\
\hline
\end{tabular}
\label{tab:tableB1}
\end{center}
\end{table*}

We present the SHMR for each satellite bin in Fig.~\ref{fig:illustris300_shmr}. The solid red circles linked by a dashed line represent the fiducial results, which show that the SHMR increases with projected physical cluster-centric radius, from $M_{\rm enh}/M_*=4.87_{-1.08}^{+1.15}$ at $R_{\rm pp} =0.13~{\rm Mpc/h}$, to $76.78^{+36.54}_{-28.75}$ at $R_{\rm pp}=0.89~{\rm Mpc/h}$. This increase in SHMR reflects the significant mass loss experienced by subhalos after they fall into the host halo, likely due to tidal stripping effects.

For the inner three $R_{\rm p}$ bins, we split the satellite galaxies into High-${\rm M_*}$ (green triangles) and Low-${\rm M_*}$ (black triangles) sub-samples. See Appendix~\ref{sec:a-a} for detailed sub-sample binning. We list the best-fit model parameters for each sub-sample in Table.~\ref{tab:tableB1}, and the corresponding lensing signals are shown in Fig.~\ref{fig:Bestfit_Rp_photo_highM_lowM}. Although the subhalo masses of the High-${\rm M_*}$ subsample are systematically higher than those of the Low-${\rm M_*}$ subsample within the same $R_{\rm p}$ range, we find no significant difference between the two subsamples in terms of the SHMR.

In Fig.~\ref{fig:illustris300_shmr}, we have plotted the observational results from various literature sources, and our results agree with those from \citet{li2016} and \citet{Niemiec2017:1703.03348v2}, where a trend of increasing SHMR with projected halo-centric radius was observed. On the other side, \citet{Sifón2015:1507.00737v3} found that SHMR has only a weak dependence on $R_{\rm pp}$ and \citet{Sifón2017:1706.06125v2} showed an anti-U shaped trend of SHMR-$R_{\rm pp}$. It should be noted that the redMaPPer cluster catalog, which includes only red-sequence galaxies, was used in \citet{li2016}, \citet{Niemiec2017:1703.03348v2}, and this work, whereas \citet{Sifón2015:1507.00737v3} and \citet{Sifón2017:1706.06125v2} did not restrict the color of member galaxies, and the galaxies in \citet{Sifón2017:1706.06125v2} have a much smaller mean stellar mass than those used in our study. However, it is unclear whether these differences in galaxy selection can account for the discrepancies shown in Fig.~\ref{fig:illustris300_shmr}.

In Fig.~\ref{fig:illustris300_shmr}, we also compare our observational results with the theoretical predictions from the state-of-art hydrodynamical simulation, TNG300-1 of the IllustrisTNG Project \citep{Nelson_2018MNRAS.475..624N, Springel_2018MNRAS.475..676S, Pillepich_2018MNRAS.475..648P, Naiman_2018MNRAS.477.1206N, Marinacci_2018MNRAS.480.5113M, Pillepich_2019MNRAS.490.3196P, Nelson_2019ComAC...6....2N}. We choose to use TNG300-1 simulation, which has a box size of $\sim$ 300 ${\rm Mpc}^3$, a dark matter mass resolution of $5.9 \times 10^7 {\rm M_{\odot}}$ and a baryonic elements (stellar particles and gas cells) mass resolution of $1.1 \times 10^7 {\rm M_{\odot}}$, where a statistical sample of analogs of redMaPPer clusters can be found. We select red satellite galaxies in TNG300 simulation whose stellar mass is larger than $1\times10^{10}{\rm M_{\odot}/h}$ and the corresponding main-halo mass $M_{\rm 200c}$ is larger than ${\rm 1\times 10^{14}{\rm M_{\odot}/h}}$, which precisely corresponds to the selection conditions of our observation samples, i.e. $M_{*}>10^{10}~{\rm M_{\odot}/h}$, $\lambda>20$. The definition of red galaxies is $g-r>0.5$, where $g$ and $r$ are the magnitudes in the SDSS $g-$band and $r-$band of galaxies provided by TNG300. We chose to use the snapshot data at $z=0.32$ because this snapshot is closest to the average redshift of all samples. In Fig.~\ref{fig:illustris300_shmr}, the solid line presents the median value of SHMR ($\frac{M_{\rm subfind}}{M_{*}}$, $M_{\rm subfind}$ is the subfind subhalo mass), and the upper and lower boundaries of the shaded area represent the 16th and 84th percentile (i.e. the $\pm 1\sigma$ confidence intervals). For the innermost $R_{\rm p}$ bin, our SHMR measurements are consistent with that of TNG300 simulation within $1\sigma$ error. For the other sub-sample bins, our measurements of SHMR are much higher than that of the simulation. In Appendix~\ref{subsec:test_method}, we demonstrate that the fitted subhalo mass from lensing signal, $M_{\rm enh}$, can effectively represent the subfind subhalo mass with TNG300-1 simulation data.

Following \citet{Niemiec2017:1703.03348v2},  we calculate the mass loss rate of satellite galaxies as
\begin{equation}
    \tau_{\rm strip}=1-\frac{M_{\rm enh}}{M_{\rm infall}} \,,
\end{equation}
\noindent where $M_{\rm infall}$ represents the dark matter mass of the satellite galaxy before it falls into the galaxy cluster. In this project, we assume the satellite galaxies have the same SHMR as those field galaxies before they fall into the galaxy clusters. We adopt the $M_{*}$-SHMR for field galaxies derived by \citet{Shan_2017} to calculate the $M_{\rm h}$, 
\begin{equation}
{\rm lg}(f_{\rm SHMR}^{-1}(M_{\rm h}))={\rm lg}(M_1)+\beta {\rm lg}(\frac{M_{\rm *}}{M_{\rm *,0}})+\frac{(\frac{M_{\rm *}}{M_{\rm *,0}})^{\delta}}{1+(\frac{M_{\rm *}}{M_{\rm *,0}})^{-\gamma}}-\frac{1}{2}\,,
\end{equation}
\noindent where ${\rm lg}(M_1)=12.52 \pm 0.050$, ${\rm lg}(M_{*,0})=10.98 \pm 0.036$, $\beta=0.47 \pm 0.022$, $\delta=0.55 \pm 0.13$ and $\gamma=1.43 \pm 0.28$ when $0.2<z<0.4$. ${\rm lg}(M_1)=12.70 \pm 0.057$, ${\rm lg}(M_{*,0})=11.11 \pm 0.038$, $\beta=0.50 \pm 0.025$, $\delta=0.54 \pm 0.16$ and $\gamma=1.72 \pm 0.30$ when $0.4<z<0.6$.
In the left panel of Fig.~\ref{fig:dm_loss_rate}, we can see that the dark matter loss rate increases with decreasing projected cluster-centric distance of the satellite galaxies. The mass loss rate of satellite galaxy subhalos shows a clear dependence on their stellar mass. This difference becomes more pronounced at larger halo-centric radii. At a projection halo-centric radius of $0.5R_{\rm 200c}$, the lower-mass subsample does not exhibit significant mass loss, while the higher-mass subsample has already lost over 80\% of its subhalo mass. However, at a projection halo-centric radius of $0.1R_{\rm 200c}$, both subsamples of satellite galaxies have lost over 80\% of their mass, with the higher-mass subsample experiencing a mass loss of over 90\%. Interestingly, the final SHMR does not exhibit a clear dependence on the stellar mass of the satellite galaxies (Fig. \ref{fig:illustris300_shmr}).

One caveat is that we assume the stellar mass remains unchanged for the satellite galaxies as they spiral into the center of the cluster. \citet{Smith2016ApJ...833..109S} studied the co-evolution of dark matter and stars in satellite galaxies and found that the stars lose about 10\% of their mass when 80\% dark matter lost. If we take this effect into account, the satellite galaxies at the center of the clusters should be compared with field galaxies with higher stellar mass, and as a result, these satellites should have an even higher mass loss rate than presented here.

We compare the retain dark matter mass fraction $M_{\rm enh}/M_{\rm infall}$ with predictions from simulations in the right panel of Fig.~\ref{fig:dm_loss_rate}. The red, green, and black lines represent the same sub-samples as in the left panel. The orange circles with error bars represent the results from \citet{xielizhi_2015MNRAS.454.1697X} with the Phoenix simulation \citep{Gao_2012_Phoenix_MNRAS.425.2169G}. The solid orange circles represent the retained mass fraction of subhaloes with the present subhalo mass $M_{\rm subfind}$ to host halo mass $M_{\rm h}$ ratio ranging from $1 \times 10^{-6}$ to $1 \times 10^{-5}$ as a function of cluster-centric distance, while the empty circles represent the results for subhaloes with $M_{\rm subfind}/M_{\rm h} > 1 \times 10^{-5}$. We also plot theoretical predictions of \citet{Han2016} using the \textsc{SubGen} code\footnote{\url{http://kambrian.github.io/SubGen/}}. We generated theoretical predictions for a galaxy cluster with $M_{\rm 200c}=2.39 \times 10^{14}~{\rm M_{\odot}/h}$, which is the average mass of our whole sample, along with the evolution of its subhalos. Subhalos are massive than $10^{-6}M_{\rm 200c}$ at the infall time. We select the satellite galaxies in this simulated galaxy cluster with $M_{*}>1\times 10^{10}~{\rm M_{\odot}/h}$. The median, 16th, and 84th percentiles of the retained dark matter mass fraction of selected satellite galaxies are represented by the solid pink line and the dashed pink lines, respectively. The observed trend of retained mass fraction as a function of halo-centric radius is broadly consistent with theoretical expectations. However, in the innermost region of galaxy clusters, the observed retained mass is lower compared to the predictions of the Phoenix Cluster simulations, but it is in better agreement with \citet{Han2016}. On the outskirts of galaxy clusters, the observed retained mass is similar to that from Phoenix Cluster but significantly higher than \citet{Han2016}. The results suggest that future studies should include hydrodynamical simulations for comparison to better understand the discrepancies between observations and theory, as well as their implications for the process of galaxy formation.

In previous figures, we bin the satellite galaxies according to their projected halo-centric distances. In this project, we also try to stack satellite galaxies of all $R_{\rm p}$, while binning the sample according to their stellar mass as shown in Appendix~\ref{sec:a-b}. The lensing signal and the best-fit model for each of these five sub-samples are shown in Fig.~\ref{fig:Mstar_Bins_ggl_best_model}. The average $R_{\rm p}$ value of five stellar mass bins are similar, with values of 0.33, 0.34, 0.36, 0.38, and 0.39 cMpc/h, respectively. In Fig.~\ref{fig:averaged_striping_rate}, we plot the average stellar mass versus their subhalo mass in the left panel. The red solid line represents the function obtained by \citet{Niemiec_2019MNRAS.487..653N} with satellite galaxies at redshift $z=0.35$ in the Illustris-1 simulation. The brown solid line represents the best-fit model for the stellar mass and dark matter mass of satellite galaxies at $z=0.24$ in TNG300, as fitted by \citet{Niemiec2022MNRAS.512.6021N}. The green solid line corresponds to the fitted relationship between the stellar and dark matter masses for central/field galaxies \citep{Shan_2017}. The orange (blue) solid line shows the relation between stellar mass and dark halo mass of satellite galaxies with weak gravitational lensing~\citep{Dvornik_2020A&A...642A..83D}. In the right panel, we show the dark matter strip rate versus stellar mass with black solid circles with error bars. The average stripping rate is lowest for satellite galaxies of $\sim 4 \times 10^{10}~{\rm M_{\odot}/h}$ with $\tau_{\rm strip}=0.59^{+0.10}_{-0.12}$, and increase to $\tau_{\rm strip}=0.91_{-0.02}^{+0.02}$ for the most massive bin of $\left<M_*\right> \sim 1.5 \times 10^{11}~{\rm M_{\odot}/h}$. The orange solid circles represent the strip rate of satellite galaxies in Illustris-1 with stellar masses between $2 \times 10^{7}$ and $2 \times 10^{11}~{\rm M_{\odot}/h}$, and the horizontal gray line shows the average strip rate of satellite galaxies in Illustris-1 calculated by \citet{Niemiec_2019MNRAS.487..653N}. The dark violet line shows the average dark matter stripping rate of passive satellite galaxies in TNG300 and the pink shows that of all satellite galaxies, both results come from \citet{Niemiec2022MNRAS.512.6021N}. The dark blue solid line represents the theoretical value of dark matter strip rate obtained by a theoretical model that combines the abundance matching technique with the halo occupation distribution and conditional luminosity (or stellar mass) function from \citet{2013ApJ...767...92R}. Results of \citet{Niemiec_2019MNRAS.487..653N} indicate that the average strip rate is nearly independent of the stellar mass, while the results of \citet{2013ApJ...767...92R} show a decrease in the loss of dark matter mass for larger stellar mass, which is opposite to our observation results.

\section{Summary and Conclusions}
\label{sec:summary_and_conclusions}
In this paper, we have performed galaxy-galaxy lensing analysis for satellite galaxies in redMaPPer galaxy clusters, derived the subhalo mass of these satellite galaxies as a function of projected halo-centric radius, and calculated the mass stripping rate of satellite galaxies. We obtain  the following conclusions.

(1) We find  $M_{\rm enh}/M_*$ decreases significantly with decreasing projected halo-centric radius, reaching $4.87^{+1.15}_{-1.08}$ at $R_{\rm pp}=0.13~{\rm Mpc/h}$, indicating dramatic mass loss due to stripping of the host halo. Our results at confirm conclusions from previous measurements of redMaPPer cluster satellite galaxy samples and galaxy-galaxy lensing \citep{li2016, Niemiec2017:1703.03348v2} at a higher S/N (see Fig.~\ref{fig:illustris300_shmr}).

(2) We provide the first measurement of the variation of dark matter mass loss rate as a function of projected halo-centric distance.  Previously, this variation could only be obtained through simulations or abundance matching.  We find satellite galaxies with larger stellar masses lose more dark matter and have higher dark matter strip rates at the same projected radius. The difference in dark matter strip rates between High-${\rm M_{*}}$ and Low-${\rm M_{*}}$ sub-samples decreases as $R_{\rm pp}$ decreases. At positions very close to the cluster center ($\sim 0.1 \times R_{\rm 200c}$), the dark matter mass loss rate for all satellite galaxies reaches $\sim 80\%$. On the other hand, the SHMR of satellite galaxies does not depend on the stellar mass of the satellite galaxies (see Fig.~\ref{fig:dm_loss_rate})

(3) We find that the average dark matter stripping rate for satellite galaxies is approximately $\sim 73\%$. The stripping rate is lowest for satellite galaxies with $\left<M_*\right> \sim 4 \times 10^{10}~{\rm M_{\odot}/h}$ and increases with $M_*$ for more massive satellite galaxies, reaching $\sim 91\%$ for satellite galaxies with $\left<M_*\right> \sim 1.5 \times 10^{11}~{\rm M_{\odot}/h}$. While our results broadly agree with the theoretical predictions from the Illustris-1 simulation, we reveal a variation of the stripping rate as a function of stellar mass, which is not seen in the simulation (see Fig.~\ref{fig:averaged_striping_rate}).

These results demonstrate that satellite galaxy-galaxy lensing is a crucial tool to understand the co-evolution of galaxies and dark matter halos. The next generation of galaxy surveys, such as the Euclid~\citep{Euclid2011arXiv1110.3193L}, the Vera Rubin Legacy Survey of Space and Time~\citep[LSST;][]{LSST_2019}~\citep{LSST_2019}, and the China Space Station Telescope~\citep[CSST;][]{zhanhu2011SSPMA..41.1441Z, Zhanhu2021}, will provide one order of magnitude larger samples of background galaxies suitable for weak lensing analysis than the current DECals survey. These upcoming surveys will allow us to more accurately measure the evolution of satellite subhalo properties in various dark matter halos.

\section{ACKNOWLEDGEMENTS}
 We acknowledge the support by National Key R\&D Program of China No. 2022YFF0503403, the support of National Nature Science Foundation of China (Nos 11988101,12022306), the support from the Ministry of Science and Technology of China (Nos. 2020SKA0110100), the science research grants from the China Manned Space Project (Nos. CMS-CSST-2021-B01, CMS-CSST-2021-A01), CAS Project for Young Scientists in Basic Research (No. YSBR-062), and the support from K.C.Wong Education Foundation. HYS acknowledges the support from NSFC of China under grant 11973070, Key Research Program of Frontier Sciences, CAS, Grant No. ZDBS-LY-7013 and Program of Shanghai Academic/Technology Research Leader. We acknowledge the support from the science research grants from the China Manned Space Project with NO. CMS-CSST-2021-A01, CMS-CSST-2021-A04. WWX acknowledges support from the National Science Foundation of China (11721303, 11890693, 12203063) and the National Key R$\&$D Program of China (2016YFA0400703). JY acknowledges the support from NSFC Grant No. 12203084, the China Postdoctoral Science Foundation Grant No. 2021T140451, and the Shanghai Post-doctoral Excellence Program Grant No. 2021419.

\section{Data Availability}

The data underlying this article will be shared on a reasonable request to the authors.

%The inclusion of a Data Availability Statement is a requirement for articles published in MNRAS. Data Availability Statements provide a standardised format for readers to understand the availability of data underlying the research results described in the article. The statement may refer to original data generated in the course of the study or to third-party data analysed in the article. The statement should describe and provide means of access, where possible, by linking to the data or providing the required accession numbers for the relevant databases or DOIs.

%%%%%%%%%%%%%%%%%%%% REFERENCES %%%%%%%%%%%%%%%%%%
% The best way to enter references is to use BibTeX:
\bibliographystyle{mnras}
\bibliography{mnras} % if your bibtex file is called example.bib
\appendix
\section{}
\label{subsec:test_method}

\begin{figure}
\centering
\includegraphics[width=0.48\textwidth]{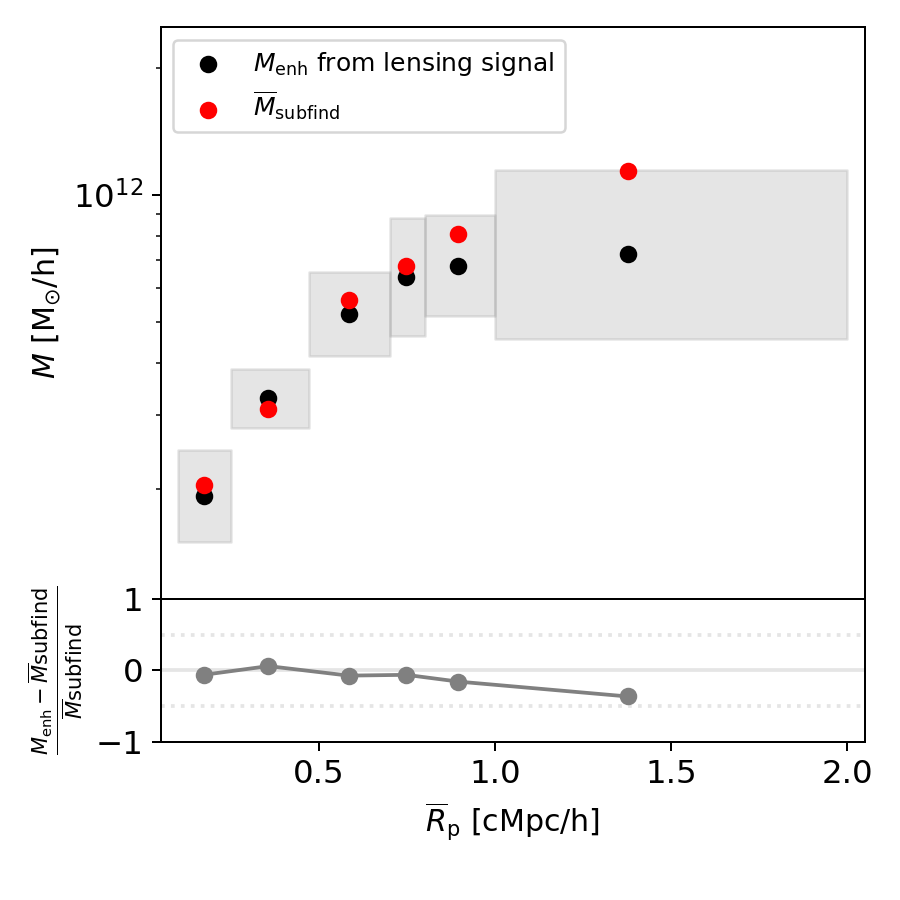}
\caption{Comparison between subhalo mass $M_{\rm enh}$ derived from lensing signals and the average value of subfind mass, $\overline{M}_{\rm subfind}$. In the upper subplot, black solid circles represent subhalo mass $M_{\rm enh}$ derived from lensing signals, while red solid circles indicate $\overline{M}_{\rm subfind}$. The gray shaded area represents the $1\sigma$ error of $M_{\rm enh}$,  which is estimated from the relative error of subhalo mass obtained by fitting the real observational data from the corresponding sub-samples. The lower subplot illustrates the variation of $\frac{M_{\rm enh}-\overline{M}_{\rm subfind}}{\overline{M}_{\rm subfind}}$ with the averaged projected halo-centric radius $\overline{R}_{\rm p}$. The solid gray line represents $y=0$. The two grey dotted lines represent $y=0.5$ and $y=-0.5$.}
\label{fig:check_tng_Rp}
\end{figure}

\begin{figure}
\centering
\includegraphics[width=0.48\textwidth]{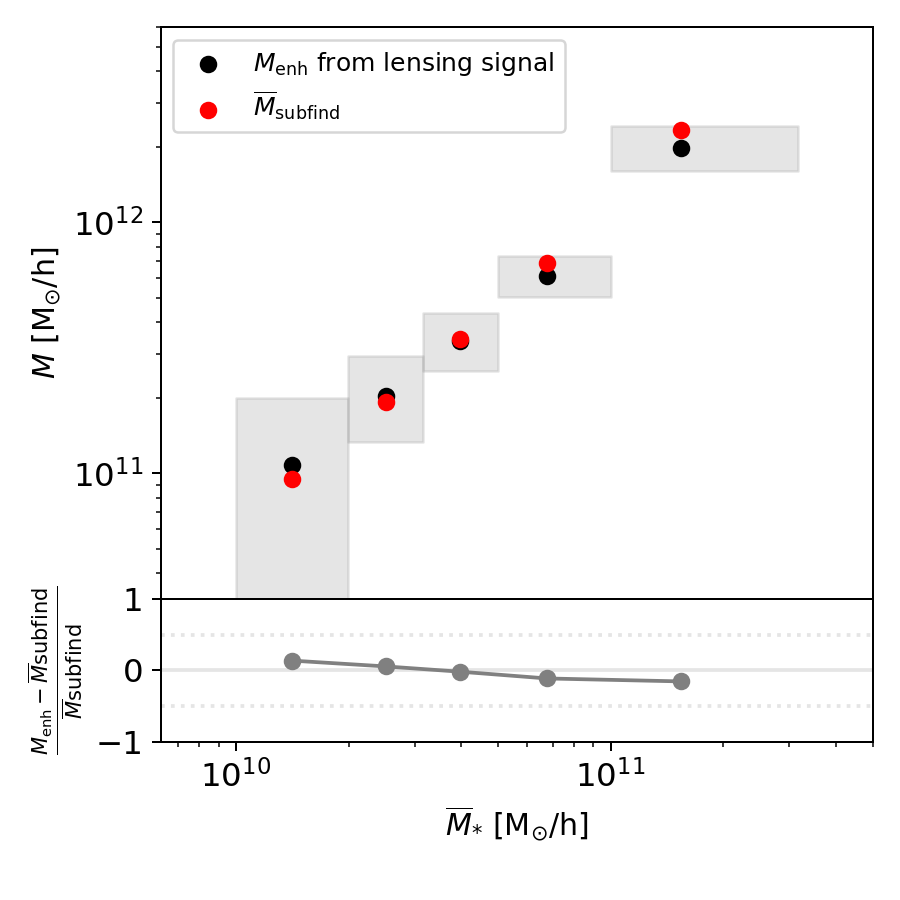}
\caption{Similar to Fig.~\ref{fig:check_tng_Rp}, this figure shows the results of sub-samples binned by stellar mass $M_{*}$. The horizontal axis represents the average stellar mass of sub-samples.}
\label{fig:check_tng_Mstar}
\end{figure}

To validate our method with simulation data, we selected red satellite galaxies with $g-r>0.5$, $M_{*}>1\times10^{10}{\rm M_{\odot}/h}$ and the corresponding main-halo mass $M_{\rm 200c}$ is larger than ${\rm 1\times 10^{14}{\rm M_{\odot}/h}}$ from the TNG300-1 simulation at $z=0.32$. We binned the satellite galaxies based on their projected halo-centric radius $R_{\rm p}$ on the $x-y$ plane or by their stellar mass $M_{*}$. By stacking the satellite-central galaxy pairs, we obtained the excess surface density $\Delta\Sigma(R)$, and subsequently, we derived subhalo mass $M_{\rm enh}$ by fitting this gravitational lensing signal with the same method as we did for the obsevational lensing signal. We compare $M_{\rm enh}$ with the corresponding average subfind mass $M_{\rm subfind}$ of each sub-sample. The comparison results of $R_{\rm p}$ binned sub-samples and $M_{*}$ binned sub-samples are presented in Figs.~\ref{fig:check_tng_Rp} and \ref{fig:check_tng_Mstar}. In both figures, black dots represent the subhalo mass, $M_{\rm enh}$, obtained from fitting the lensing signal, while the red solid dots represent the corresponding average subfind subhalo mass $\overline{M}_{\rm subfind}$. The gray shaded area represents the $1\sigma$ confidence interval of $M_{\rm enh}$, which is estimated from the relative error of subhalo mass obtained by fitting the real observational data from the corresponding sub-samples. As we can see that $M_{\rm enh}$ and $\overline{M}_{\rm subfind}$ are consistent within $1\sigma$ confidence interval and the relative deviations $\frac{M_{\rm enh}-\overline{M}_{\rm subfind}}{\overline{M}_{\rm subfind}}$ are The relative deviation are within $\pm$16\% for the vast majority of sub-samples, indicating that the fitted subhalo mass from lensing signal can effectively represent the subfind subhalo mass.

\section{}
\label{sec:a-a}
\begin{figure*}
\centering
\includegraphics[width=0.4\textwidth]{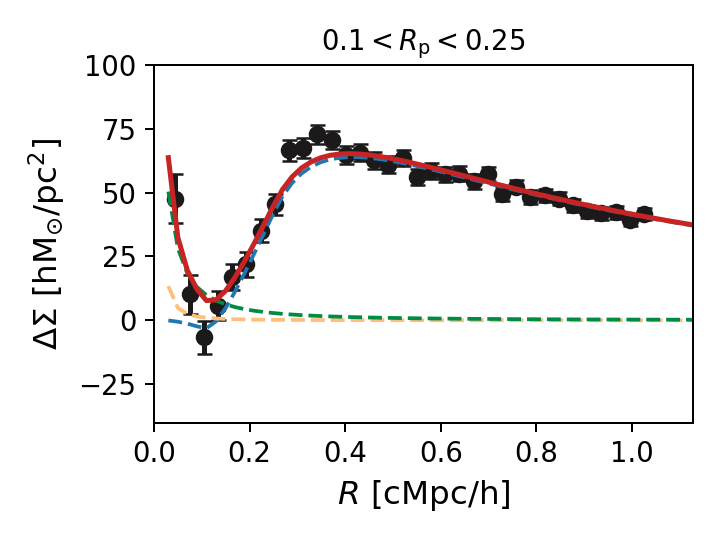}
\includegraphics[width=0.4\textwidth]{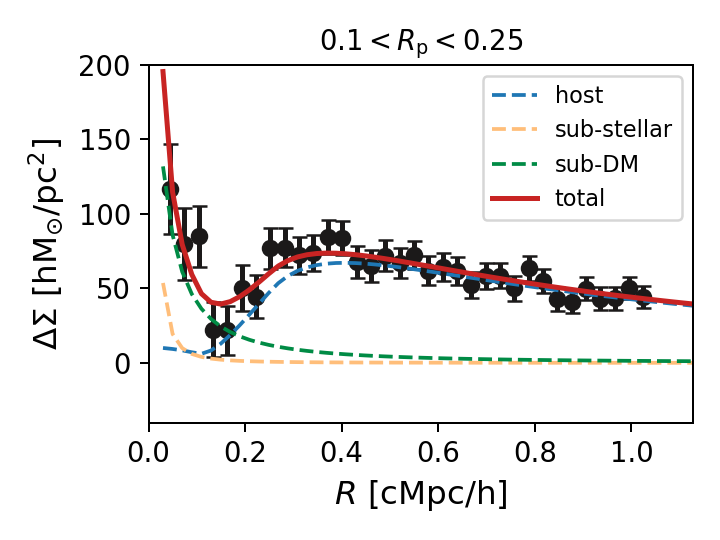}

\includegraphics[width=0.4\textwidth]{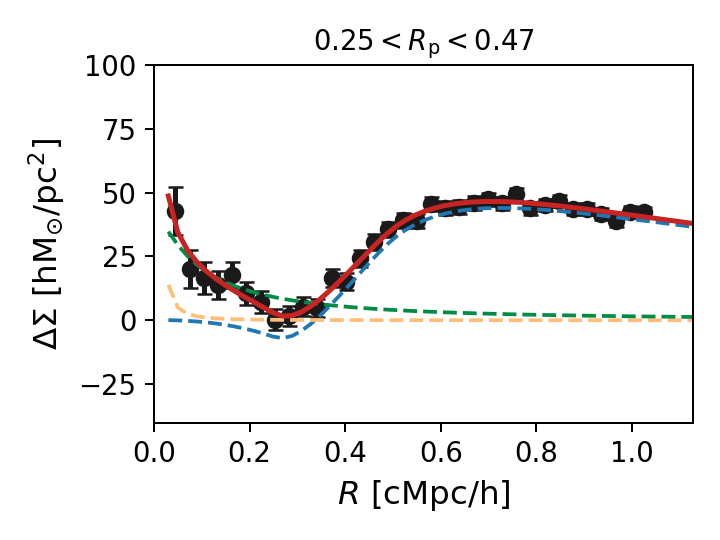}
\includegraphics[width=0.4\textwidth]{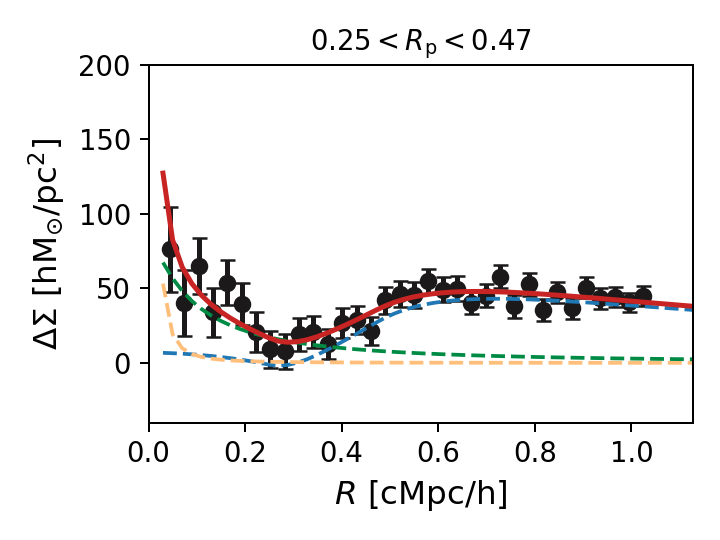}

\includegraphics[width=0.4\textwidth]{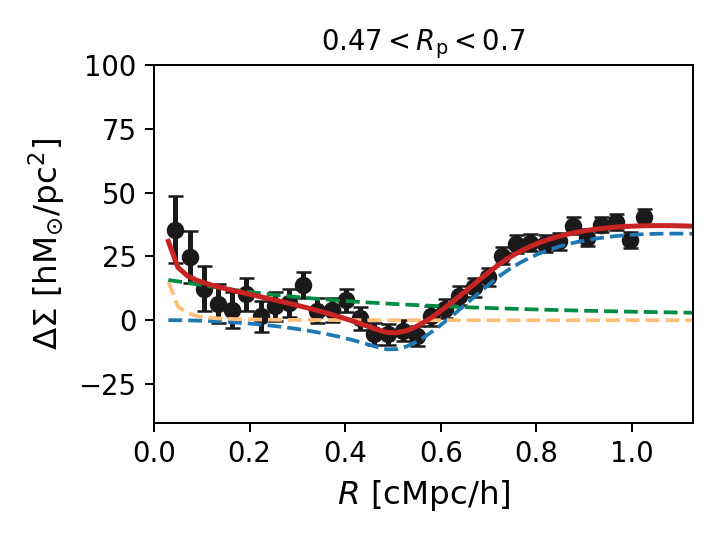}
\includegraphics[width=0.4\textwidth]{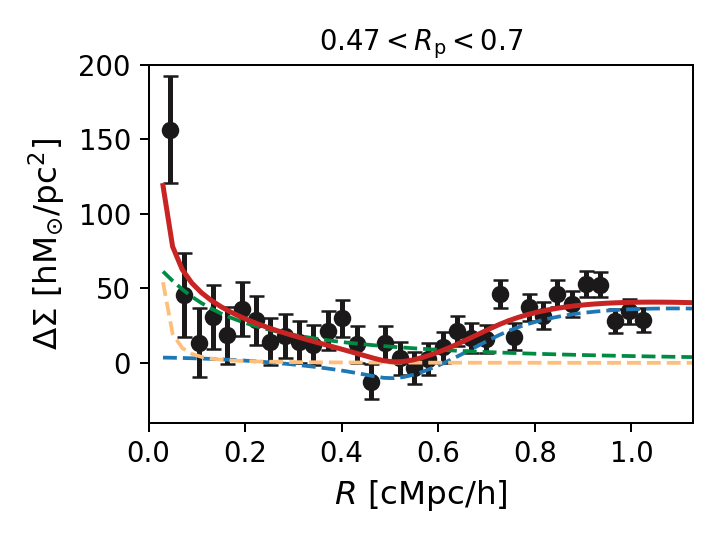}

\caption{Similar to Fig.~\ref{fig:Bestfit_Rp_photo}, but here we show the lensing signals and best-fit models corresponding to the Low-${\rm M}_{*}$ (left column) and High-${\rm M}_{*}$ (right column) sub-samples.}
\label{fig:Bestfit_Rp_photo_highM_lowM}
\end{figure*}

To test whether the SHMR depends on stellar mass, we divide each of the smallest three $R_{\rm p}$ sub-sample in Sec.~\ref{subsec:lens} into two sub-samples, namely High-${\rm M}_{*}$ ($10^{11}{\rm M_{\odot}/h} <M_{*}< 10^{12}{\rm M_{\odot}/h}$) and Low-${\rm M}_{*}$ ($10^{10}{\rm M_{\odot}/h} <M_{*}< 10^{11}{\rm M_{\odot}/h}$) sub-samples. Here we present the gravitational lensing signals and the best-fit model of the High-${\rm M}_{*}$ and Low-${\rm M}_{*}$ sub-samples in Fig.~\ref{fig:Bestfit_Rp_photo_highM_lowM}. The number of lenses and best-fit parameters of each sub-sample are listed in Table.~\ref{tab:table1}.

\section{}
\label{sec:a-b}
\begin{figure*}
\includegraphics[width=0.98\textwidth]{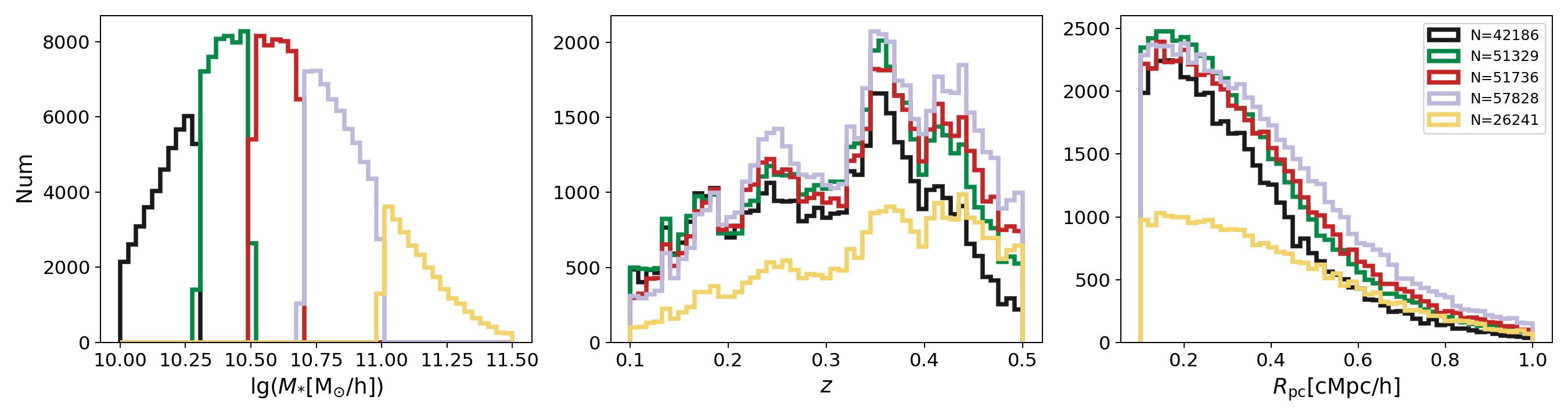}
\caption{Similar to Fig.~\ref{fig:6bins}, here we show the histogram distributions of stellar mass $M_{*}$, redshift $z$, and comoving lensing distance $R_{\rm p}$ for sub-samples binned solely based on $M_{*}$. The same sub-samples are represented with consistent colors across the three panels.}
\label{fig:averaged_5bins}
\end{figure*}

\begin{figure*}
\centering
\includegraphics[width=0.4\textwidth]{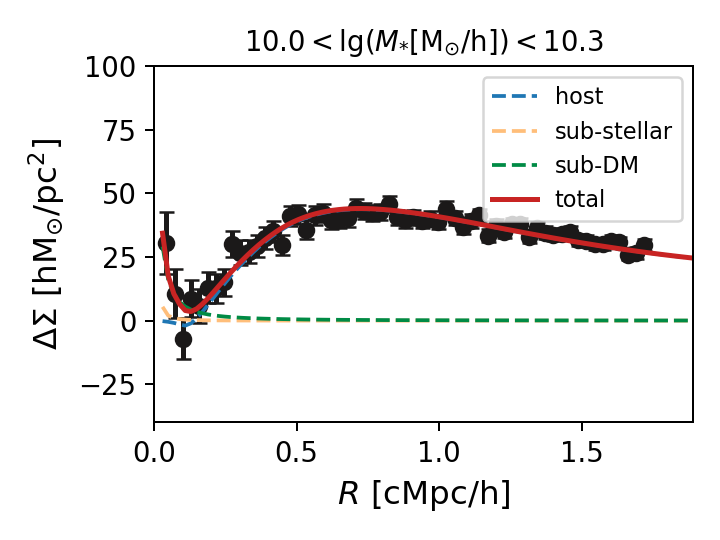}
\includegraphics[width=0.4\textwidth]{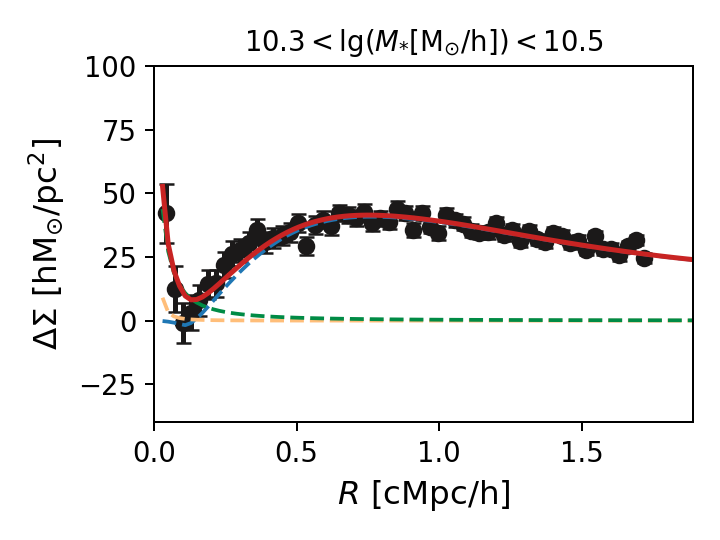}
\includegraphics[width=0.4\textwidth]{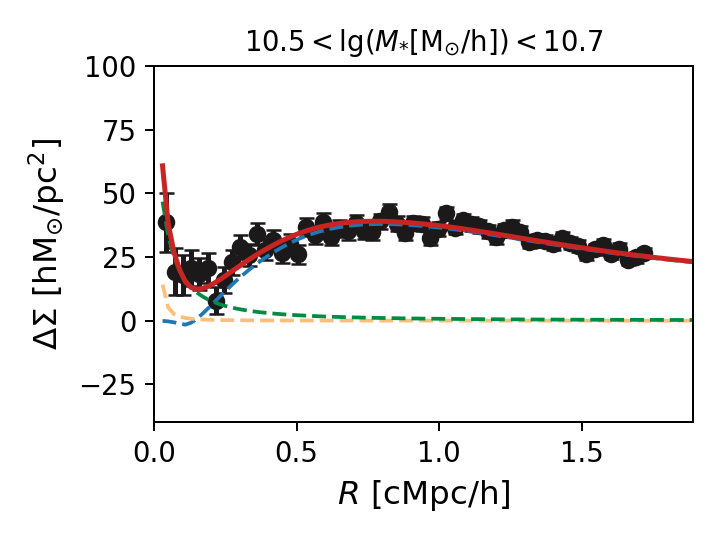}
\includegraphics[width=0.4\textwidth]{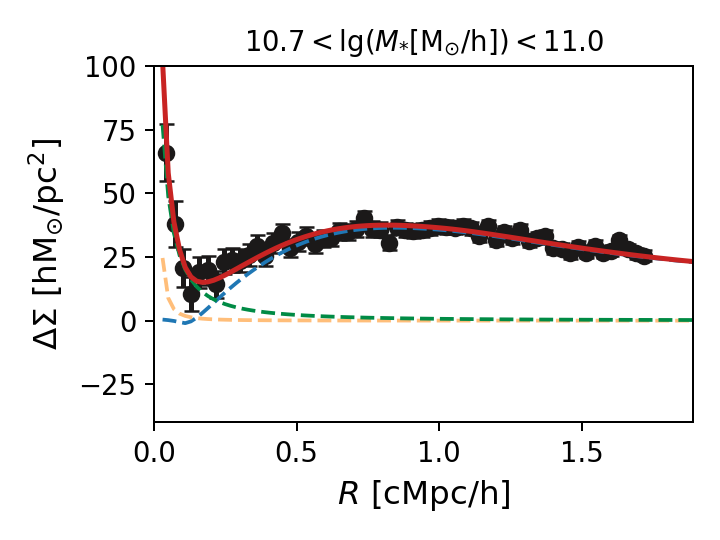}
\includegraphics[width=0.4\textwidth]{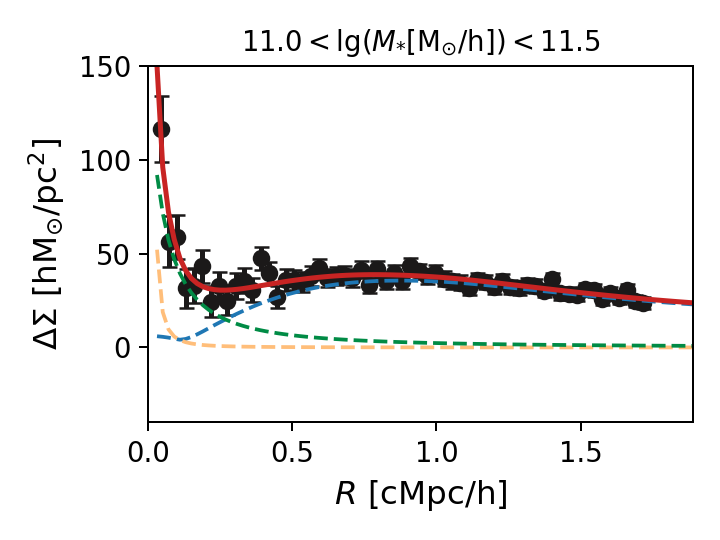}
\caption{Similar to Fig.~\ref{fig:Bestfit_Rp_photo}, here we show the lensing signal and the best-fit model of sub-samples binned solely based on $M_{*}$.}
\label{fig:Mstar_Bins_ggl_best_model}
\end{figure*}

To obtain the average dark matter stripping rate of satellite galaxies in different stellar mass ranges, we divided the sample with stellar masses ranging from $10^{10}~{\rm M_{\odot}/h}$ to $10^{11.5}~{\rm M_{\odot}/h}$ and satisfying the criteria of $0.1<z<0.5$, $P_{\rm mem}>0.8$, ${\rm 0.1~cMpc/h}<R_{\rm p}<{\rm 1.0~cMpc/h}$, and ${\rm DEC}<34$ into five sub-samples. The distributions of stellar mass, redshift, and comoving lensing distance to the central galaxy for each subsample are shown in Fig.~\ref{fig:averaged_5bins}, with the same color used to represent the same subsample in all three panels. The five sub-samples have very similar redshift distributions, and the $R_{\rm p}$ distributions of the four lower stellar mass bins are also very similar. However, the sub-sample with the largest stellar mass has a relatively larger $R_{\rm p}$ projection distance. The bin edges of the stellar mass and the corresponding number of satellite galaxies in each bin, as well as the best-fit model parameters, are listed in Table.~\ref{tab:tableB1}. 

We also use the SWOT software to calculate the lensing signals for different sub-samples (60 linear radial bins, $0.05~{\rm cMpc/h}<R<1.75~{\rm cMpc/h}$), and fit the lensing signals with MCMC sampler \textsc{Emcee}. The lensing signals and best-fit models of different sub-samples are shown in Fig.~\ref{fig:Mstar_Bins_ggl_best_model}.

% Don't change these lines
\bsp	% typesetting comment
\label{lastpage}
\end{document}